\documentclass[aps,prd,epsfig,preprint,amsmath,amssymb,superscriptaddress,nofootinbib]{revtex4}
\usepackage{multirow}
\usepackage{graphicx,bm}
\usepackage[usenames]{color}
\usepackage{float}
\usepackage{caption}
\usepackage{subcaption}
\usepackage{slashed}
\begin{document}
\draft
\title{The effects of Higgs boson couplings through $HZZ$ production at future lepton colliders}

\author{Serdar Spor}
\email[]{serdar.spor@beun.edu.tr}
\affiliation{Department of Medical Imaging Techniques, Zonguldak B\"{u}lent Ecevit University, 67100, Zonguldak, T\"{u}rkiye.}

\begin{abstract}

We focus on the sensitivity of the anomalous Higgs-gauge boson couplings at $H\gamma Z$ and $HZZ$ vertices through the process $\ell^- \ell^+ \rightarrow HZZ$ at CLIC and Muon Collider. Signal and relevant backgrounds events are generated in MadGraph within Standard Model Effective Field Theory (SMEFT) framework. These events are passed through Pythia for parton showering, and realistic detector effects are simulated by Delphes. The limits at 95\% C.L. on the coefficients $\overline{c}_{HB}$ and $\overline{c}_{HW}$ are obtained at two b-tagging working points; loose and medium containing the Delphes card from CLIC and Muon Collider, corresponding to b-tagging efficiencies of 90\% and 70\%, respectively. We report that our best 95\% C.L. limits on $\overline{c}_{HB}$ and $\overline{c}_{HW}$ coefficients are $[-0.00138; 0.00090]$ and $[-0.00162; 0.00026]$, respectively, at 3 TeV CLIC with an integrated luminosity of 5 ab$^{-1}$, and $[-0.00024; 0.00023]$ and $[-0.00020; 0.00009]$, respectively, at 10 TeV Muon Collider with an integrated luminosity of 10 ab$^{-1}$. These limits are compared with the present experimental and various phenomenological limits.

\end{abstract}


\maketitle

\section{Introduction} \label{Sec1}

The discovery of the Higgs boson by the ATLAS \cite{Aad:2012les} and CMS \cite{Chatrchyan:2012les} collaborations at the Large Hadron Collider (LHC) has been a fundamental milestone in confirming the electroweak symmetry breaking mechanism predicted by the Standard Model (SM). Although the SM is largely consistent with experimental measurements to date, it cannot answer many fundamental questions. This suggests that new physics effects may arise beyond the SM, making precise investigation of the Higgs boson's interactions particularly important. The deviations in the measured properties of the Higgs boson may be indirect indicators of new physics scenarios that are not yet directly observable. In this context, a detailed examination of the Higgs boson's interactions with electroweak gauge bosons offers a powerful tool for testing the limits of the SM. Independently of the model, the most commonly used approach to describe such potential deviations is Standard Model Effective Field Theory (SMEFT). Within the SMEFT framework, anomalous Higgs-gauge boson interactions (such as $H\gamma\gamma$, $HZ\gamma$, $HZZ$, and $HWW$) are parametrically investigated via higher-dimensional operators, and deviations from SM predictions are revealed, expressing new physics effects.

Future colliders offer unique opportunities to investigate such anomalous coupling interactions. Future colliders with multi-TeV energy and high integrated luminosity make precise measurements of rare processes and multiple boson final states possible. However, lepton colliders offer significant advantages over hadron colliders, providing a clean experimental environment for studying anomalous Higgs interactions. Since leptons are elementary particles, much better control is achieved over the energy of colliding particles. Unlike hadron colliders, where large QCD backgrounds and proton remnants, lepton colliders provide well-defined initial states, virtually no pile-up, and negligible QCD activity.

We consider two types of future lepton collider: the Compact Linear Collider (CLIC) for  electron-positron collision, and the Muon Collider for muon-antimuon collision. The parameters for the stage of these two colliders are as follows: $\sqrt{s}=3$ TeV and ${\cal L}_{\text{int}}=5$ ab$^{-1}$ for CLIC \cite{Robson:2018tgb} and $\sqrt{s}=10$ TeV and ${\cal L}_{\text{int}}=10$ ab$^{-1}$ for Muon Collider \cite{Aime:2022tqz}. Significant evaluations and research are underway regarding their construction. This study demonstrates that both colliders could be useful for investigating the effects of the anomalous Higgs interactions under consideration, focusing on examining deviations from the SM in the interactions of the Higgs boson with gauge bosons. Within the framework of the SMEFT approach, the effects of anomalous Higgs interactions are studied through processes $e^-e^+/\mu^-\mu^+ \rightarrow HZZ$ at CLIC and Muon Collider, and the sensitivity limits are evaluated. The aim is to contribute to the discovery of possible new physics signatures in the Higgs sector based on the results obtained. Recently, many studies have been conducted on the determination of anomalous Higgs boson couplings in CLIC \cite{Abramowicz:2017wxz,Ellis:2017pmw,Denizli:2018rca,Sahin:2019unz,Karadeniz:2020yvz,Roloff:2020efg,Spor:2024ghq,Lima:2025kdw} and Muon Collider \cite{Chiesa:2020yhn,Han:2021pas,Chen:2021pln,Han:2021pmq,Chen:2022ygc,Forslund:2022wzc,Blas:2022tzs,Spor:2024yln,Spor:2025eaz,Gurkanli:2025wcv}.

\section{Anomalous Higgs-gauge boson interactions} \label{Sec2}

In particle physics, the SM successfully describes fundamental particles and their interactions as a quantum field theory defined under $SU(3)_C \times SU(2)_L \times U(1)_Y$ gauge symmetry. However, the SM is insufficient for directly describing new physics effects. Therefore, the Effective Field Theory (EFT) approach is widely used to study the indirect effects of physics beyond SM. The EFT offers a model-independent framework that bridges the experimentally accessible electroweak energy scale with new physics effects expected to emerge at higher energy scales. In this approach, new physics effects are parameterized through high-dimensional operators added to the SM Lagrangian. The validity of EFT depends on the condition that the new physics scale $\Lambda$ is much larger than the typical energy scale $E$ of the process of
interest $(\Lambda \gg E)$. When this condition is met, the EFT offers a reliable approximation. Within the framework of SMEFT, high-dimensional operators consisting of SM fields and respecting SM gauge symmetries are used. The most dominant contributions generally come from dimension-six operators, while dimension-eight and higher operators are neglected due to their additional suppression. In this approach, lepton and baryon number conservation is ensured, and new physics effects are studied as free parameters via Wilson coefficients.

The EFT framework is particularly effective for investigating the properties and interactions of the Higgs boson in a model-independent way. In this regard, the Strongly Interacting Light Higgs (SILH) basis offers an important theoretical structure. The SILH basis assumes the Higgs boson as a pseudo-Goldstone boson arising from a strongly interacting sector and predicts that electroweak symmetry breaking occurs through this mechanism \cite{Tosciri:2021yhb}. The effective Lagrangian is defined by including the four-dimensional SM Lagrangian and six-dimensional operators as follows \cite{Giudice:2007ops,Alloul:2014hws}:

\begin{eqnarray}
\label{eq.1} 
{\cal L}_{\text{eff}}={\cal L}_{\text{SM}}+\sum_{i}{\overline{c}_i}{\cal O}_i+\sum_{i}{\widetilde{c}_i}{\cal O}_i
\end{eqnarray}

{\raggedright where ${\cal O}_i$ are the dimension-six operators, ${\overline{c}_i}$ and ${\widetilde{c}_i}$ are the CP-conserving and CP-violating Wilson coefficients, respectively. The SILH Lagrangian with the CP-conserving operators is written below \cite{Giudice:2007ops,Alloul:2014hws}}

\begin{eqnarray}
\label{eq.2} 
\begin{split}
{\cal L}_{\text{SILH}}=&\frac{\overline{c}_H}{2\upsilon^2}\partial^\mu (\Phi^\dagger \Phi)\partial_\mu(\Phi^\dagger \Phi)+\frac{\overline{c}_T}{2\upsilon^2}(\Phi^\dagger \overset\leftrightarrow{D}^\mu \Phi)(\Phi^\dagger \overset\leftrightarrow{D}_\mu \Phi) \\
&+\frac{ig\overline{c}_W}{m_W^2}(\Phi^\dagger T_{2k}\overset\leftrightarrow{D}^\mu\Phi)D^\nu W_{\mu\nu}^k+\frac{ig^\prime \overline{c}_B}{2m_W^2}(\Phi^\dagger \overset\leftrightarrow{D}^\mu\Phi) \partial^\nu B_{\mu\nu} \\
&+\frac{2ig\overline{c}_{HW}}{m_W^2}(D^\mu \Phi^\dagger T_{2k}D^\nu\Phi)W_{\mu\nu}^k+\frac{ig^\prime \overline{c}_{HB}}{m_W^2}(D^\mu \Phi^\dagger D^\nu\Phi)B_{\mu\nu} \\
&+\frac{g^{\prime2}\overline{c}_{\gamma}}{m_W^2} \Phi^\dagger \Phi B_{\mu\nu} B^{\mu\nu}
\end{split}
\end{eqnarray}

{\raggedright where $\upsilon$ is the vacuum expectation value of the Higgs field. $T_{2k}=\sigma_k/2$, where $\sigma_k$ is the Pauli matrices. ${B}_{\mu\nu}=\partial_\mu B_\nu - \partial_\nu B_\mu$ and ${W}_{\mu\nu}^k=\partial_\mu W_\nu^k - \partial_\nu W_\mu^k + g\epsilon_{ij}^k W_\mu^i W_\nu^j$ are the field strength tensors. $D^\mu$ is the covariant derivative operator with $\Phi^\dagger\overset\leftrightarrow{D}_\mu\Phi=\Phi^\dagger(D_\mu \Phi)-(D_\mu \Phi^\dagger)\Phi$, where $\Phi$ is the Higgs doublet in SM. The Lagrangian of Eq.~(\ref{eq.2}) can be completed with additional CP-violating operators,}

\begin{eqnarray}
\label{eq.3} 
{\cal L}_{\text{CPV}}=\frac{ig\widetilde{c}_{HW}}{m_W^2}D^\mu \Phi^\dagger T_{2k} D^\nu \Phi \widetilde{W}^k_{\mu\nu}+\frac{ig^\prime \widetilde{c}_{HB}}{m_W^2}D^\mu \Phi^\dagger D^\nu \Phi \widetilde{B}_{\mu\nu}+\frac{g^{\prime 2}\widetilde{c}_\gamma}{m_W^2}\Phi^\dagger \Phi B_{\mu\nu}\widetilde{B}^{\mu\nu}
\end{eqnarray}

The Lagrangian containing the Wilson coefficients in the SILH basis of dimension-six CP-conserving and CP-violating operators can be defined in terms of mass eigenstates for the anomalous $H\gamma Z$ and $HZZ$ couplings after electroweak symmetry breaking as follows \cite{Alloul:2014hws}:

\begin{eqnarray}
\label{eq.4} 
\begin{split}
{\cal L}=&-\frac{1}{2}g_{h\gamma z}^{(1)}Z_{\mu\nu}F^{\mu\nu}h-g_{h\gamma z}^{(2)}Z_{\nu}\partial_\mu F^{\mu\nu}h-\frac{1}{2}\widetilde{g}_{h\gamma z}Z_{\mu\nu}\widetilde{F}^{\mu\nu}h \\
&-\frac{1}{4}g_{hzz}^{(1)}Z_{\mu\nu}Z^{\mu\nu}h-g_{hzz}^{(2)}Z_{\nu}\partial_\mu Z^{\mu\nu}h+\frac{1}{2}g_{hzz}^{(3)}Z_{\mu}Z^{\mu}h-\frac{1}{4}\widetilde{g}_{hzz}Z_{\mu\nu}\widetilde{Z}^{\mu\nu}h 
\end{split}
\end{eqnarray}     

{\raggedright where $h$, $F_{\mu\nu}$ and $Z_{\mu\nu}$ are the Higgs boson field, the field strength tensors of photon and $Z$-boson, respectively. The relationships between the Lagrangian parameters in Eqs.~(\ref{eq.2}-\ref{eq.3}) and Eq.~(\ref{eq.4}) are shown below:}

\begin{eqnarray}
\label{eq.5} 
g_{h\gamma z}^{(1)}=\frac{gs_W}{c_Wm_W}\left[\overline{c}_{HW}-\overline{c}_{HB}+8\overline{c}_{\gamma}s_W^2\right]
\end{eqnarray}
\begin{eqnarray}
\label{eq.6} 
g_{h\gamma z}^{(2)}=\frac{gs_W}{c_Wm_W}\left[\overline{c}_{HW}-\overline{c}_{HB}-\overline{c}_{B}+\overline{c}_{W}\right]
\end{eqnarray}
\begin{eqnarray}
\label{eq.7} 
\widetilde{g}_{h\gamma z}=\frac{gs_W}{c_Wm_W}\left[\widetilde{c}_{HW}-\widetilde{c}_{HB}+8\widetilde{c}_{\gamma}s_W^2\right]
\end{eqnarray}
\begin{eqnarray}
\label{eq.8} 
g_{hzz}^{(1)}=\frac{2g}{c_W^2 m_W}\left[\overline{c}_{HB}s_W^2-4\overline{c}_{\gamma}s_W^4+c_W^2 \overline{c}_{HW}\right]
\end{eqnarray}
\begin{eqnarray}
\label{eq.9} 
g_{hzz}^{(2)}=\frac{g}{c_W^2 m_W}\left[(\overline{c}_{HW}+\overline{c}_{W})c_W^2+(\overline{c}_B+\overline{c}_{HB})s_W^2\right]
\end{eqnarray}
\begin{eqnarray}
\label{eq.10} 
g_{hzz}^{(3)}=\frac{gm_W}{c_W^2}\left[1-\frac{1}{2}\overline{c}_{H}-2\overline{c}_{T}+8\overline{c}_\gamma \frac{s_W^4}{c_W^2}\right]
\end{eqnarray}
\begin{eqnarray}
\label{eq.11} 
\widetilde{g}_{hzz}=\frac{2g}{c_W^2 m_W}\left[\widetilde{c}_{HB}s_W^2-4\widetilde{c}_{\gamma}s_W^4+c_W^2\widetilde{c}_{HW}\right]
\end{eqnarray}

Here, $c_W=\text{cos}\theta_W$, $s_W=\text{sin}\theta_W$ and $\theta_W$ is the weak mixing angle. As seen in Eqs.~(\ref{eq.5}-\ref{eq.11}), the anomalous Higgs-gauge boson couplings in the effective Lagrangian for the anomalous $H\gamma Z$ and $HZZ$ couplings are sensitive to the ten Wilson coefficients; $\overline{c}_\gamma$, $\overline{c}_B$, $\overline{c}_W$, $\overline{c}_{HB}$, $\overline{c}_{HW}$, $\overline{c}_T$, $\overline{c}_H$, $\widetilde{c}_\gamma$, $\widetilde{c}_{HB}$, $\widetilde{c}_{HW}$. In this paper, the SM coupling of the Higgs boson to the $Z$ is held fixed while the anomalous couplings are varied. This corresponds to setting $\overline{c}_H$ and $\overline{c}_T$ to zero in the analysis. This turns out to have a negligible effect on the final results.

In this study, the effects of dimension-6 operators are investigated through Monte Carlo simulations performed with leading order using the {\sc MadGraph5}$\_$aMC@NLO \cite{Alwall:2014cvc} event generator. The effective Lagrangians of the SMEFT containing CP-conserving and CP-violating dimension-6 operators are implemented using the FeynRules \cite{Alloul:2014tfc} and UFO \cite{Degrande:2012opq} framework. We investigate the potential of anomalous Higgs-gauge boson couplings on the process $e^- e^+ \rightarrow HZZ \rightarrow b\bar{b}\ell \ell \nu \bar{\nu}$ at CLIC and the process $\mu^- \mu^+ \rightarrow HZZ \rightarrow b\bar{b}\ell \ell \nu \bar{\nu}$ at Muon Collider. 

\begin{figure}[H]
\centering
\begin{subfigure}{0.48\linewidth}
\includegraphics[width=\linewidth]{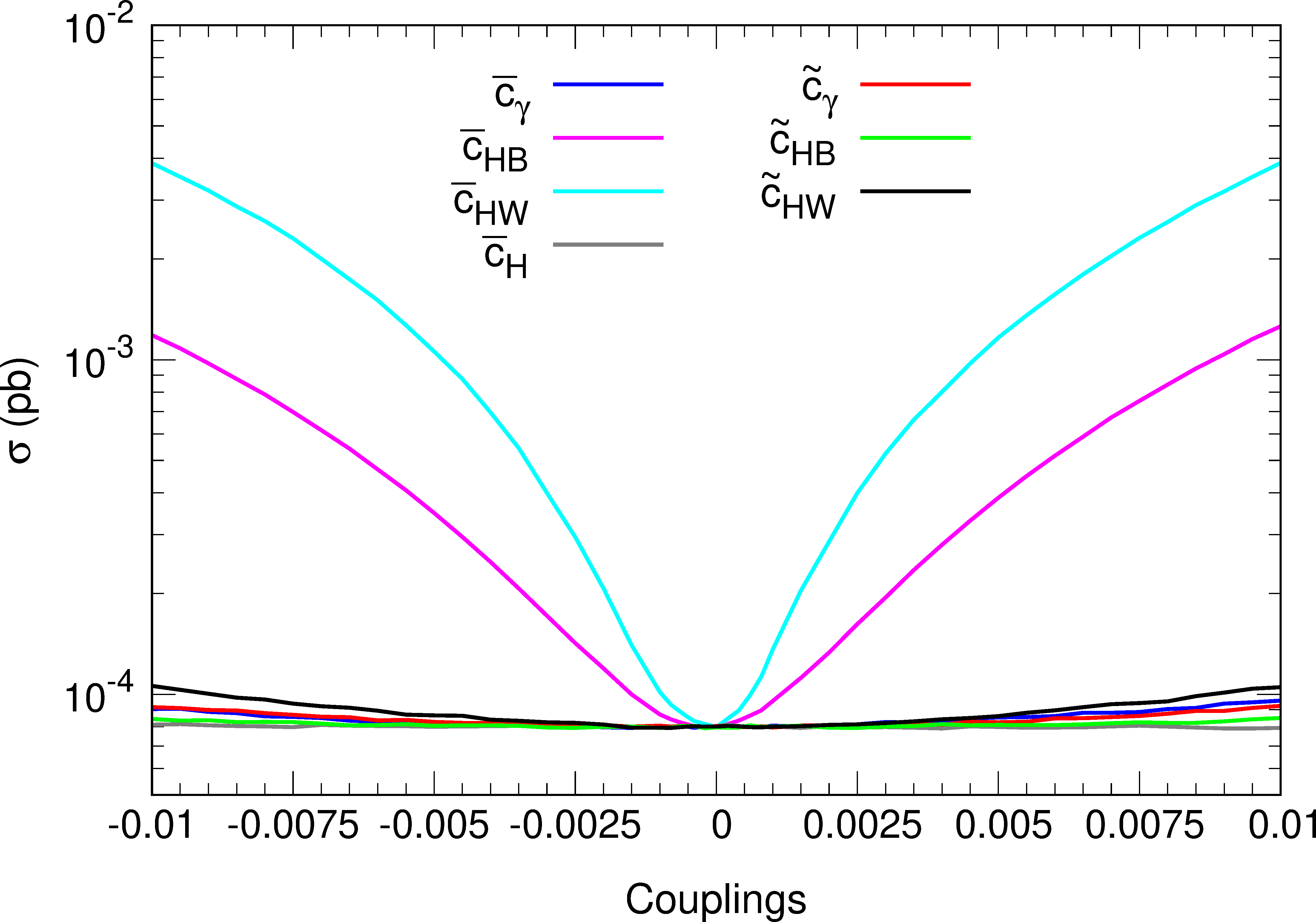}
\caption{}
\label{fig1:a}
\end{subfigure}\hfill
\begin{subfigure}{0.48\linewidth}
\includegraphics[width=\linewidth]{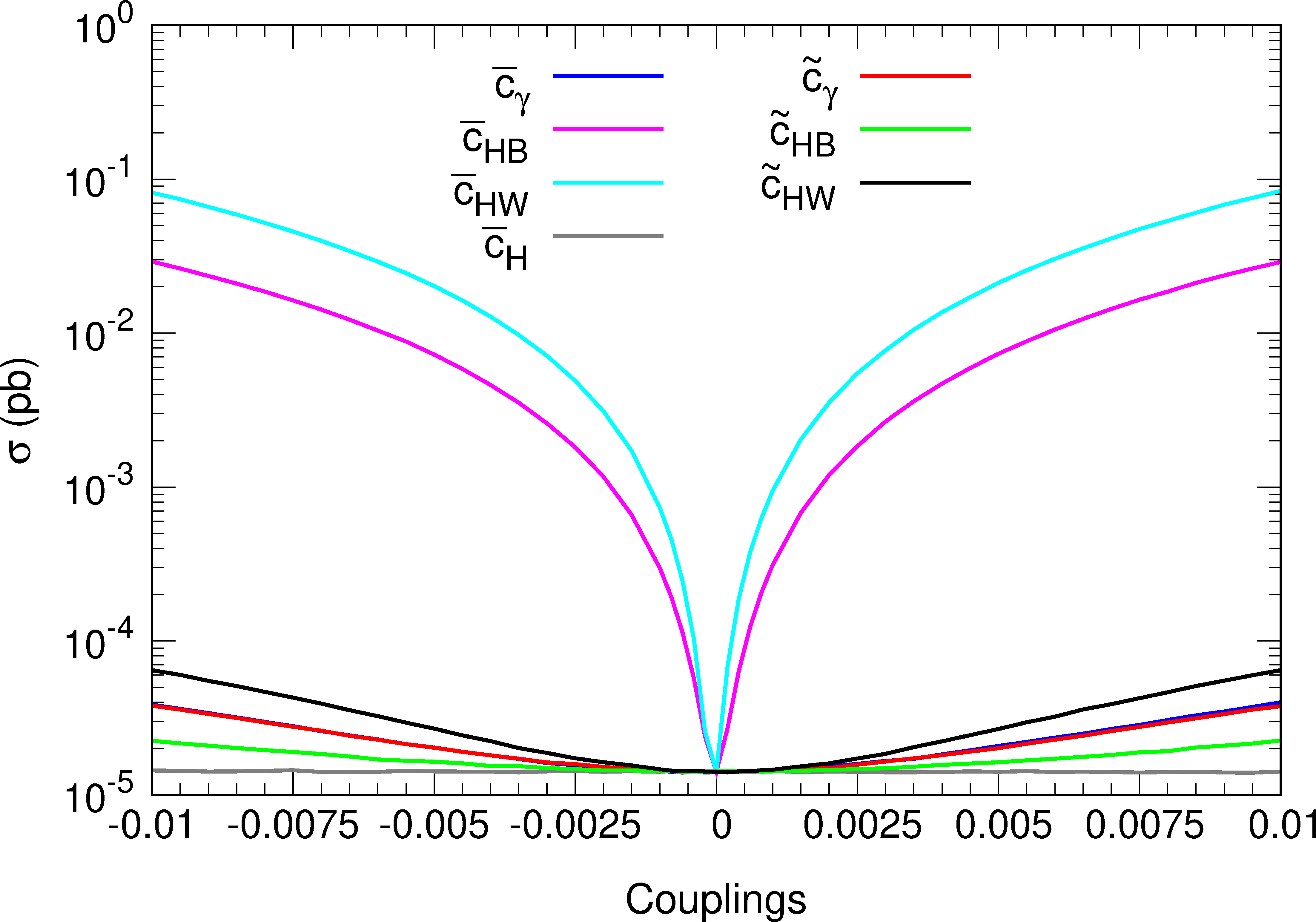}
\caption{}
\label{fig1:b}
\end{subfigure}\hfill
\caption{The total cross-section of the process $\ell- \ell+ \rightarrow HZZ$ as a function of coefficients $\overline{c}_\gamma$, $\widetilde{c}_\gamma$, $\overline{c}_H$, $\overline{c}_{HB}$, $\widetilde{c}_{HB}$, $\overline{c}_{HW}$ and $\widetilde{c}_{HW}$ at the CLIC (a) and the Muon Collider (b).}
\label{fig1}
\end{figure}

The total cross-sections of the process $\ell^- \ell^+ \rightarrow HZZ$ at generator level as a function of coefficients $\overline{c}_\gamma$, $\widetilde{c}_\gamma$, $\overline{c}_H$, $\overline{c}_{HB}$, $\widetilde{c}_{HB}$, $\overline{c}_{HW}$ and $\widetilde{c}_{HW}$ at the CLIC and the Muon Collider using the one-parameter analysis are given in Fig.~\ref{fig1:a} and Fig.~\ref{fig1:b}, respectively. The total cross-sections in Fig.~\ref{fig1} are generated by calculating a function of the coefficient in question when all other coefficients were zero. The SM cross-section corresponds to the point $\overline{c}_\gamma=\overline{c}_H=\overline{c}_{HB}=\overline{c}_{HW}=\widetilde{c}_\gamma=\widetilde{c}_{HB}=\widetilde{c}_{HW}=0$ in Fig.~\ref{fig1}. The cross-sections of $\overline{c}_{HB}$ and $\overline{c}_{HW}$ coefficients at both colliders are seen to be larger than those of the $\overline{c}_\gamma$, $\overline{c}_H$, $\widetilde{c}_\gamma$, $\widetilde{c}_{HB}$ and $\widetilde{c}_{HW}$ coefficients. Due to the greater sensitivity of $\overline{c}_{HB}$ and $\overline{c}_{HW}$ coefficients compared to the others in the $HZZ$ production process, we will focus on $\overline{c}_{HB}$ and $\overline{c}_{HW}$ coefficients in the following sections and as the aim of this study.

\section{Analysis of Signal and Background Events} \label{Sec3}

We present simulations and cut-based analysis to investigate the potential of $HZZ$ production for obtaining limits on $\overline{c}_{HB}$ and $\overline{c}_{HW}$ coefficients of the anomalous $H\gamma Z$ and $HZZ$ couplings within the model-independent SMEFT framework at CLIC and Muon Collider. The final state in the process $\ell^- \ell^+ \rightarrow HZZ \rightarrow b\bar{b}\ell \ell \nu \bar{\nu}$ is formed by the decay of Higgs boson into two $b$ quarks, and the decay of $Z$-bosons into charged lepton pair along with neutrino pair. A cut-based analysis is applied to enhance sensitivity by exploiting the expected signal kinematics.

We consider the following seven relevant backgrounds; (i) only SM contribution from the same final state of the signal process $\ell^- \ell^+ \rightarrow HZZ \rightarrow b\bar{b}\ell \ell \nu \bar{\nu}$, where the Higgs boson decays into two $b$ quarks, and the $Z$-bosons decay into charged lepton pair along with neutrino pair; (ii) the process $\ell^- \ell^+ \rightarrow ZZZ \rightarrow b\bar{b}\ell \ell \nu \bar{\nu}$, where the $Z$-bosons decay into two $b$ quarks and charged lepton pair along with neutrino pair; (iii) the total contribution of the processes $\ell^- \ell^+ \rightarrow HWW \rightarrow b\bar{b}\ell \nu \ell \bar{\nu}$ and (iv) $\ell^- \ell^+ \rightarrow ZWW \rightarrow b\bar{b}\ell \nu \ell \bar{\nu}$, where $W$-bosons decay leptonically and $H$ or $Z$-boson decay into two $b$ quarks; (v) the total of the contribution from the processes $\ell^- \ell^+ \rightarrow HZ \rightarrow b\bar{b}\ell \ell$ and (vi) $\ell^- \ell^+ \rightarrow ZZ \rightarrow b\bar{b}\ell \ell$, where the $Z$-boson decays into charged lepton pair and the $H$ or $Z$-boson decay into two $b$ quarks; (vii) the process $\ell^- \ell^+ \rightarrow t\bar{t} \rightarrow WbW\bar{b} \rightarrow \ell \nu b \ell \bar{\nu}\bar{b}$, where $W$-bosons also decay leptonically as a result of $t$ quarks decaying into $Wb$.

All signal and background events are generated 1M for each in {\sc MadGraph5}$\_$aMC@NLO and these events are processed using PYTHIA 8.3 \cite{Bierlich:2022uzx} for parton showering, hadronization, fragmentation, and decay. We use Delphes 3.5.0 \cite{Favereau:2014wfb} for a fast simulation of the detector response with delphes\_card\_CLICdet\_Stage3.tcl configuration designed for the energy stage of $\sqrt{s}=3$ TeV at CLIC and with delphes\_card\_MuonColliderDet.tcl configuration at Muon Collider. In the CLIC and Muon Collider detector cards, jet clustering is performed using the Valencia Linear Collider (VLC) algorithm \cite{Boronat:2015uev,Boronat:2018csf} in exclusive mode with a fixed number of jets $N_j = 2$ and a cone size parameter $R = 1.0$ at $\gamma=\beta=1$. The b-tagging efficiency is analyzed at two working points (WP) defined as the loose WP (90\% b-tagging efficiency) and the medium WP (70\% b-tagging efficiency). A hypothetical forward muon detector in $8.0 > |\eta| > 2.5$ includes 90\% efficiency with $0.5 < p_T <= 1$ GeV and 95\% efficiency with $p_T > 1$ GeV in the Delphes card. Misidentification rates in terms of b-tagging efficiency emerge as a function of pseudo-rapidity and energy and at the CLIC and Muon Collider with $1.53 < |\eta| \le 2.09$ and $E \ge 500$ GeV, the misidentification rates for medium WP and loose WP are 0.009 and 0.05, respectively.

Due to the short lifetime of muons, their decay products will inevitably interact with beamline components, creating numerous secondary and tertiary particles that can reach the detector. This critical experimental challenge for detector performance is called the beam-induced background (BIB). To effectively study the high-energy collisions of interest, this effect needs to be mitigated. The BIB can be partially eliminated by placing shielding nozzles near the interaction point, but this comes at the cost of losing detector acceptance up to a certain opening angle. Nevertheless, recent studies have shown encouraging signs that BIB can be successfully controlled \cite{Antonelli:2016ygs,Bartosik:2019edc,Bartosik:2020umx,Jindariani:2022wqw,Bartosik:2022ylz}. It is assumed that the impact of BIB is expected to be substantially reduced if $|\eta| \le 2.5$ and $p_T \ge 25$ GeV for all visible particles of the detector \cite{Asadi:2021uge,Homiller:2022esk}.

\subsection{CLIC Study}

\begin{figure}[H]
\centering
\begin{subfigure}{0.47\linewidth}
\includegraphics[width=\linewidth]{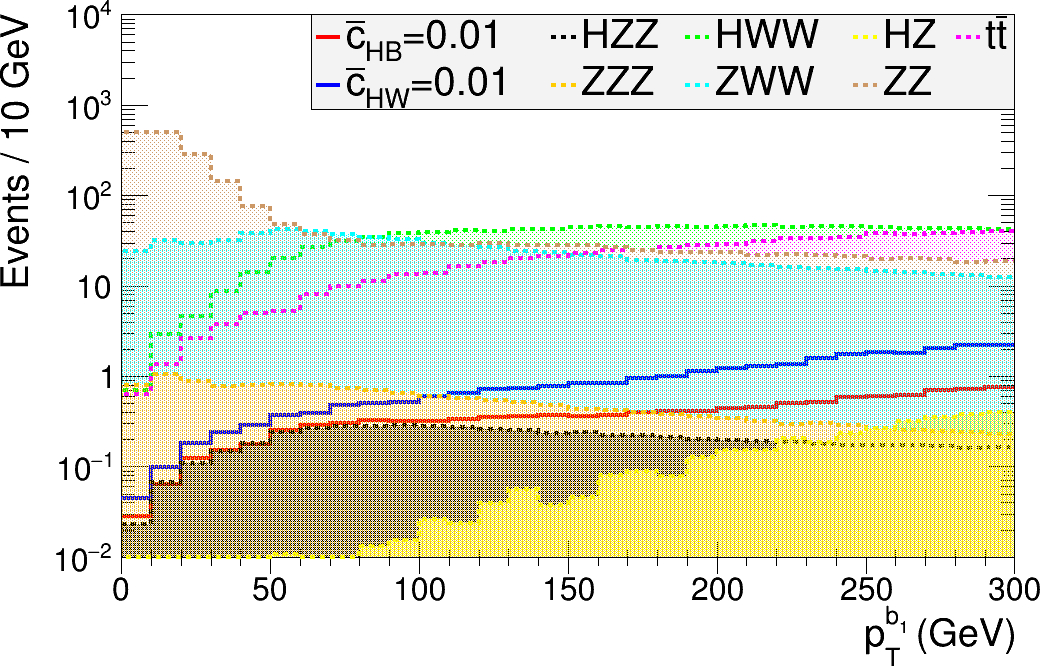}
\caption{}
\label{fig2:a}
\end{subfigure}\hfill
\begin{subfigure}{0.47\linewidth}
\includegraphics[width=\linewidth]{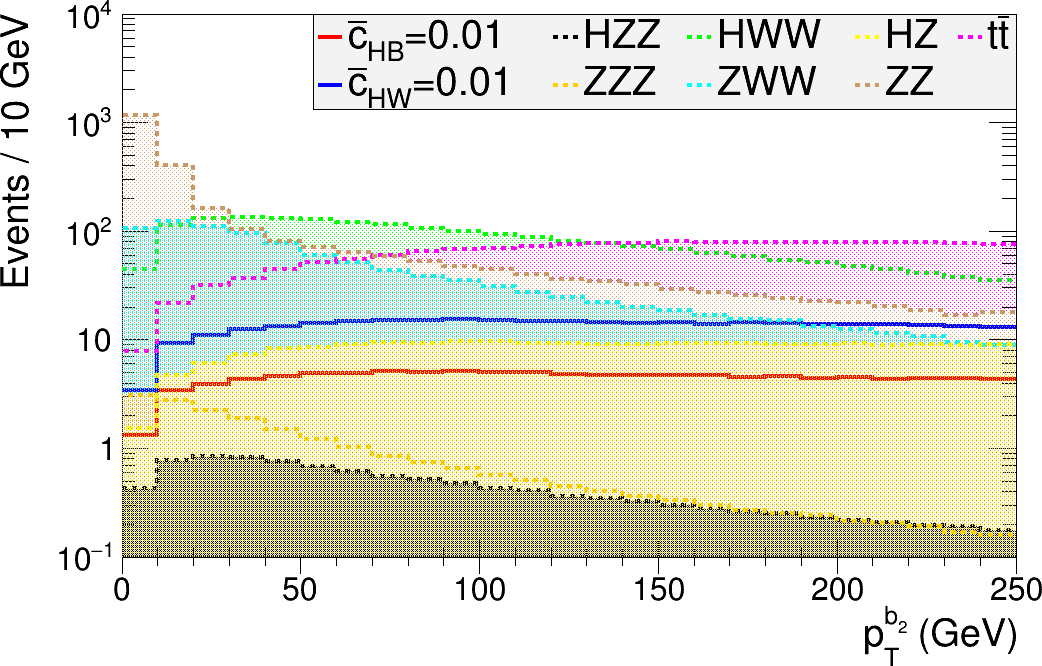}
\caption{}
\label{fig2:b}
\end{subfigure}\hfill

\begin{subfigure}{0.47\linewidth}
\includegraphics[width=\linewidth]{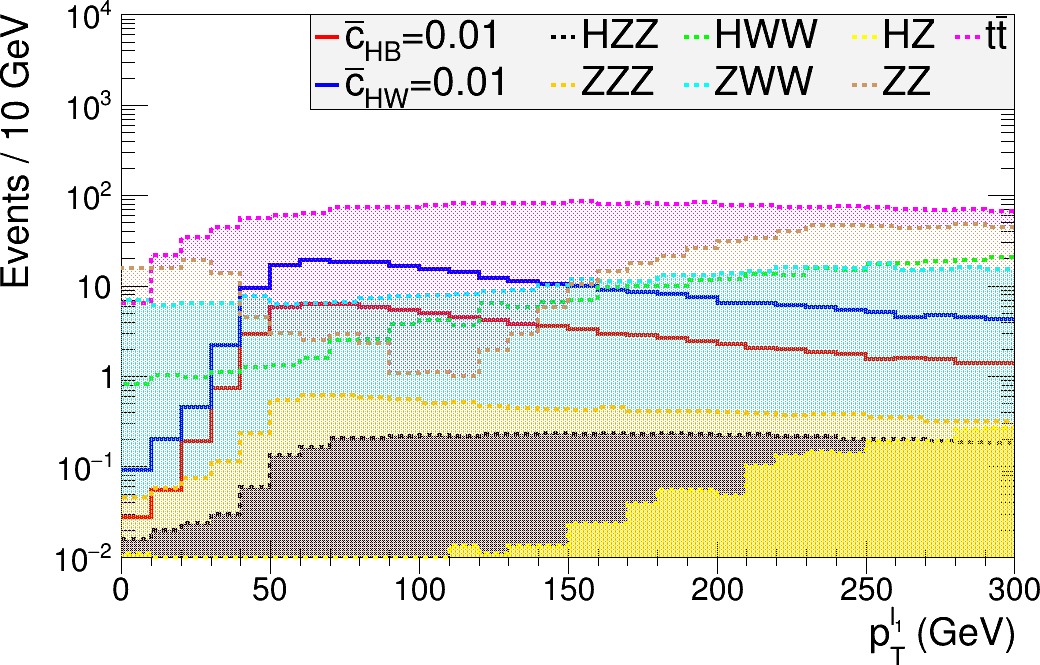}
\caption{}
\label{fig2:c}
\end{subfigure}\hfill
\begin{subfigure}{0.47\linewidth}
\includegraphics[width=\linewidth]{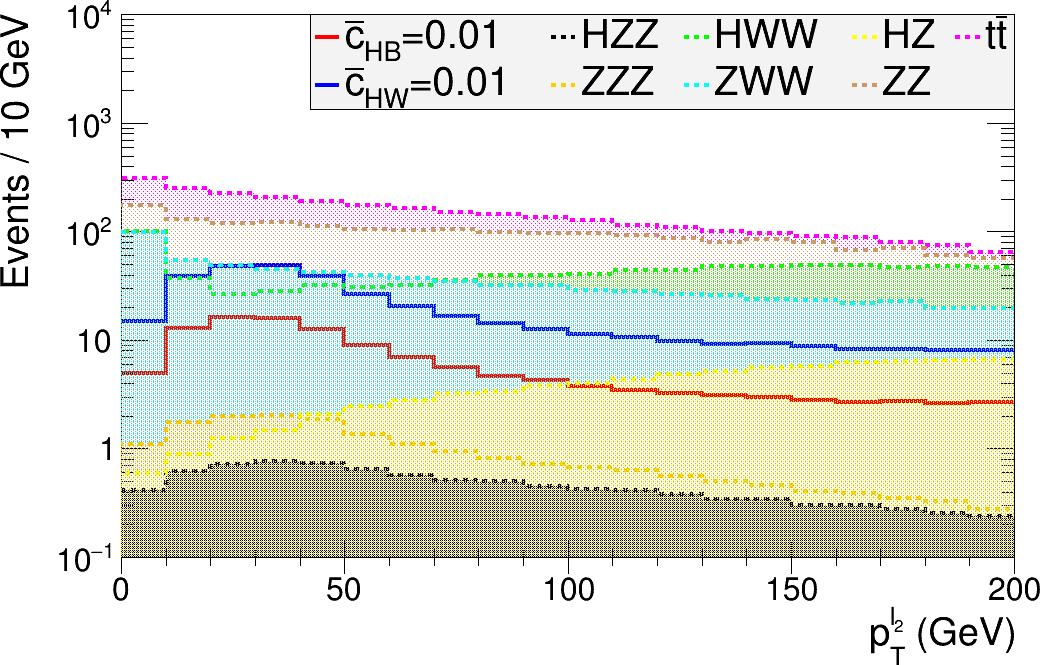}
\caption{}
\label{fig2:d}
\end{subfigure}\hfill

\caption{The distributions of transverse momentum of the leading b-tagged jet (a), the sub-leading b-tagged jet (b), the leading charged lepton (c), and the sub-leading charged lepton (d) for signal and relevant background
processes at CLIC.}
\label{fig2}
\end{figure}

\begin{figure}[H]
\centering
\begin{subfigure}{0.48\linewidth}
\includegraphics[width=\linewidth]{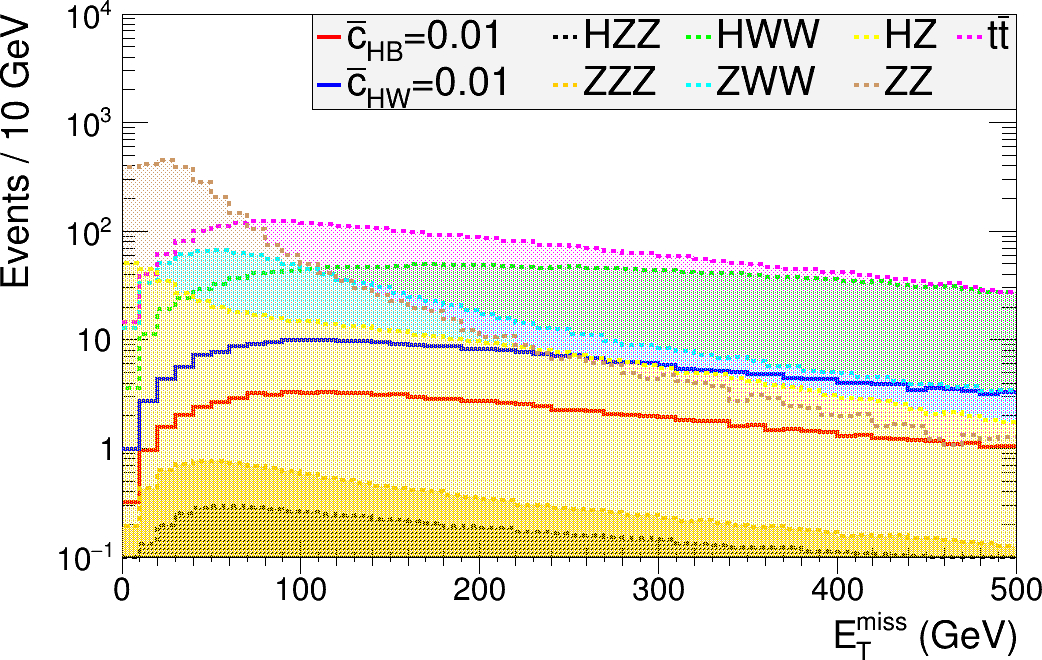}
\caption{}
\label{fig3:a}
\end{subfigure}\hfill
\begin{subfigure}{0.48\linewidth}
\includegraphics[width=\linewidth]{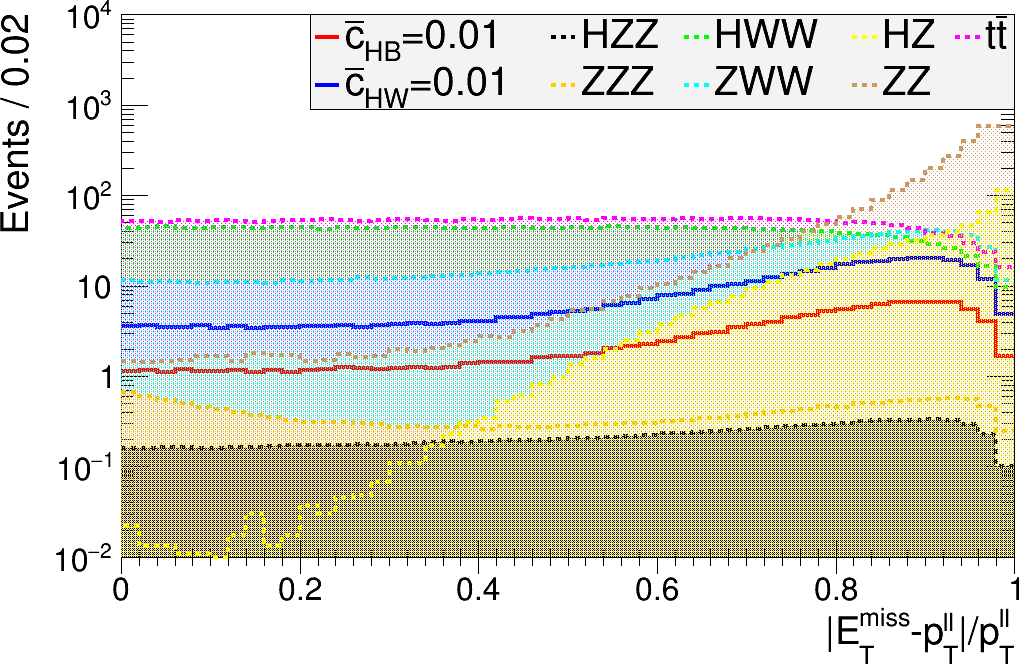}
\caption{}
\label{fig3:b}
\end{subfigure}\hfill

\begin{subfigure}{0.48\linewidth}
\includegraphics[width=\linewidth]{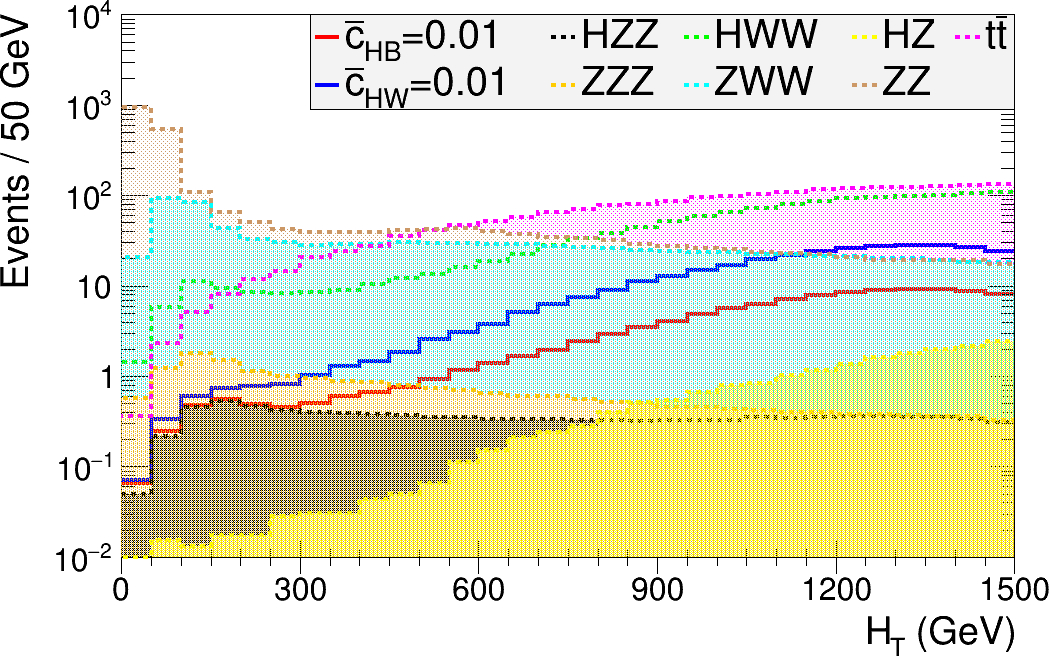}
\caption{}
\label{fig3:c}
\end{subfigure}\hfill
\begin{subfigure}{0.48\linewidth}
\includegraphics[width=\linewidth]{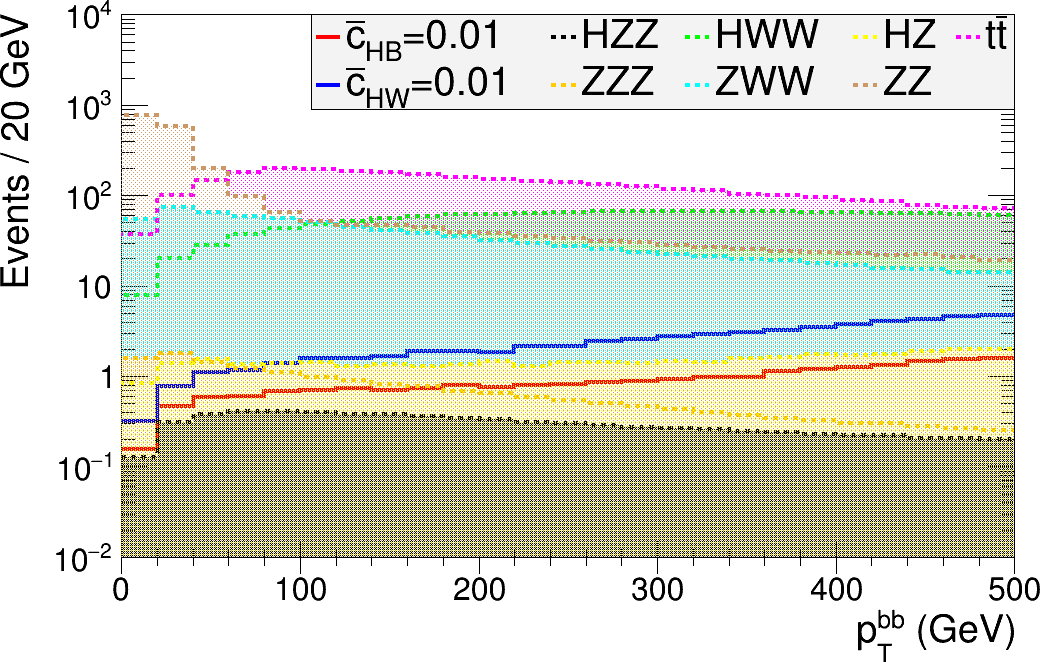}
\caption{}
\label{fig3:d}
\end{subfigure}\hfill

\caption{The distributions of the missing energy transverse (a), the transverse momentum balance ratio (b), the scalar sum of the transverse energy (c), and the transverse momentum of two b-tagged jets (d) for signal and relevant background processes at CLIC.}
\label{fig3}
\end{figure}

\begin{figure}[H]
\centering
\begin{subfigure}{0.48\linewidth}
\includegraphics[width=\linewidth]{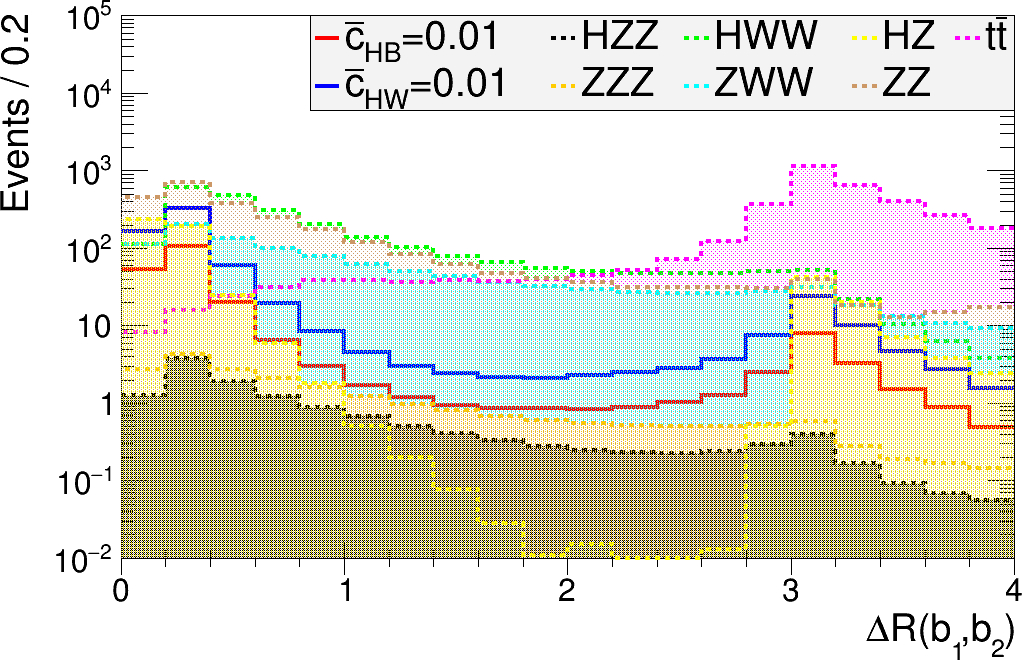}
\caption{}
\label{fig4:a}
\end{subfigure}\hfill
\begin{subfigure}{0.48\linewidth}
\includegraphics[width=\linewidth]{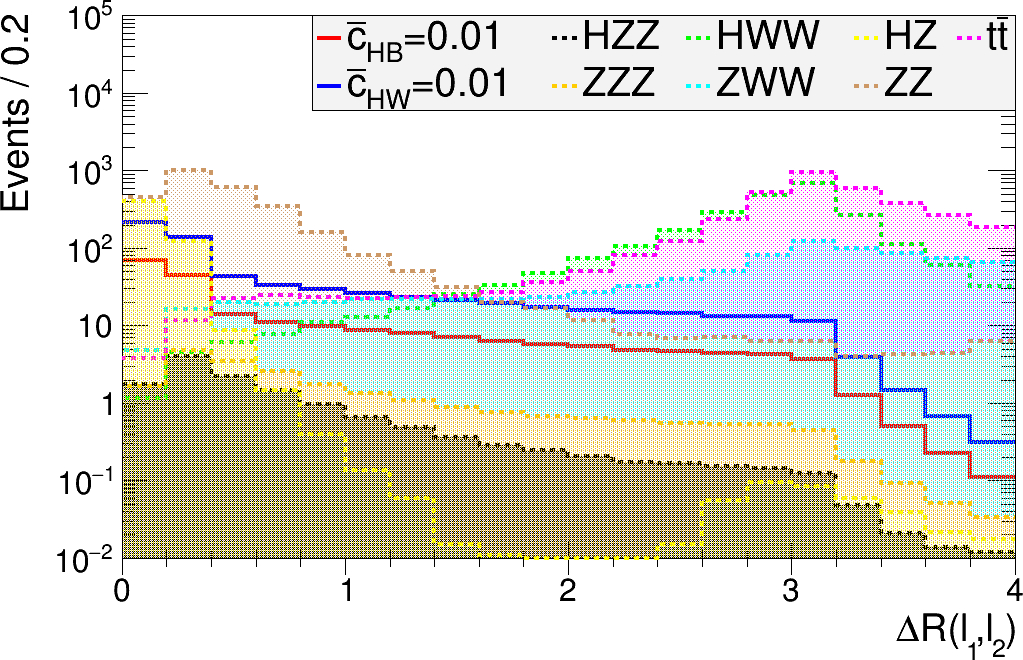}
\caption{}
\label{fig4:b}
\end{subfigure}\hfill
\begin{subfigure}{0.48\linewidth}
\includegraphics[width=\linewidth]{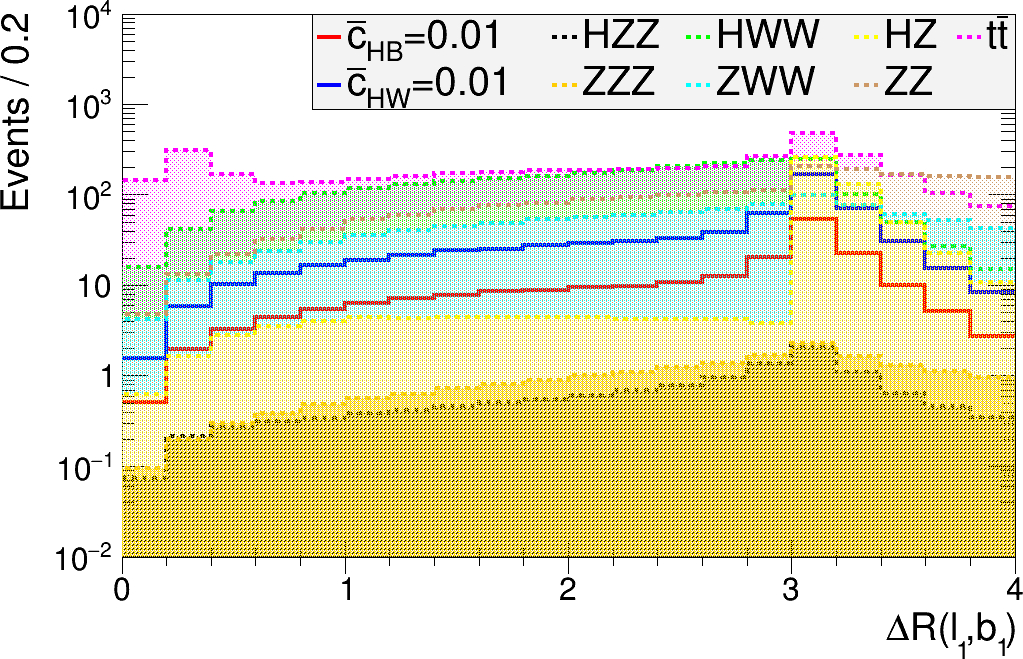}
\caption{}
\label{fig4:c}
\end{subfigure}\hfill

\caption{The distributions of the angular distance between the two b-tagged jets (a), between the two charged leptons (b), and between the leading charged lepton and the leading b-tagged jet (c) for signal and relevant background processes at CLIC.}
\label{fig4}
\end{figure}

\begin{figure}[H]
\centering
\begin{subfigure}{0.48\linewidth}
\includegraphics[width=\linewidth]{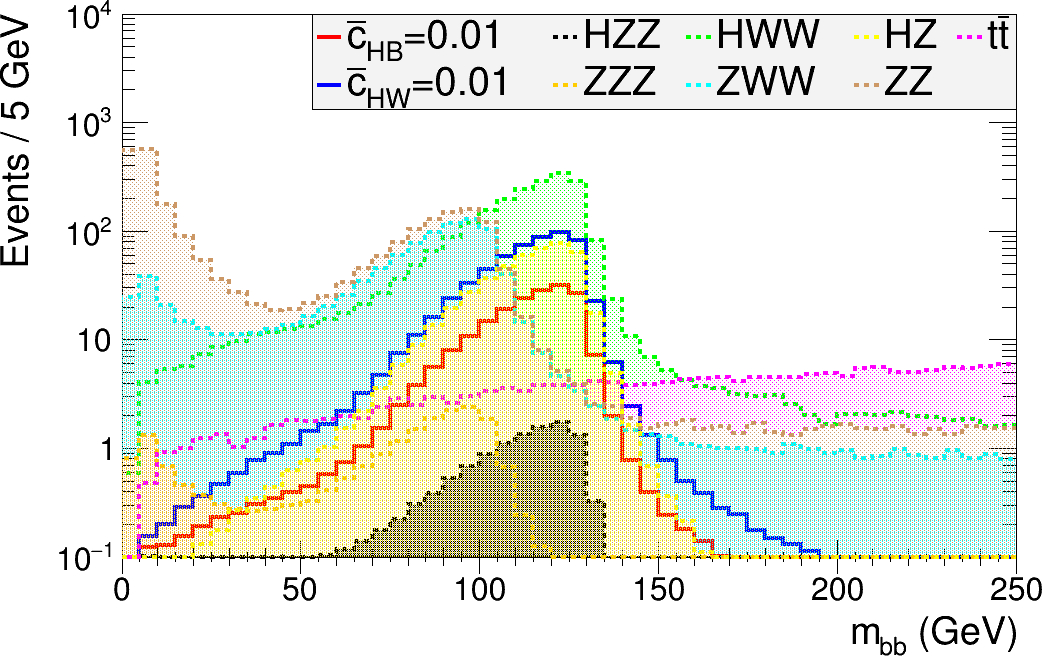}
\caption{}
\label{fig5:a}
\end{subfigure}\hfill
\begin{subfigure}{0.48\linewidth}
\includegraphics[width=\linewidth]{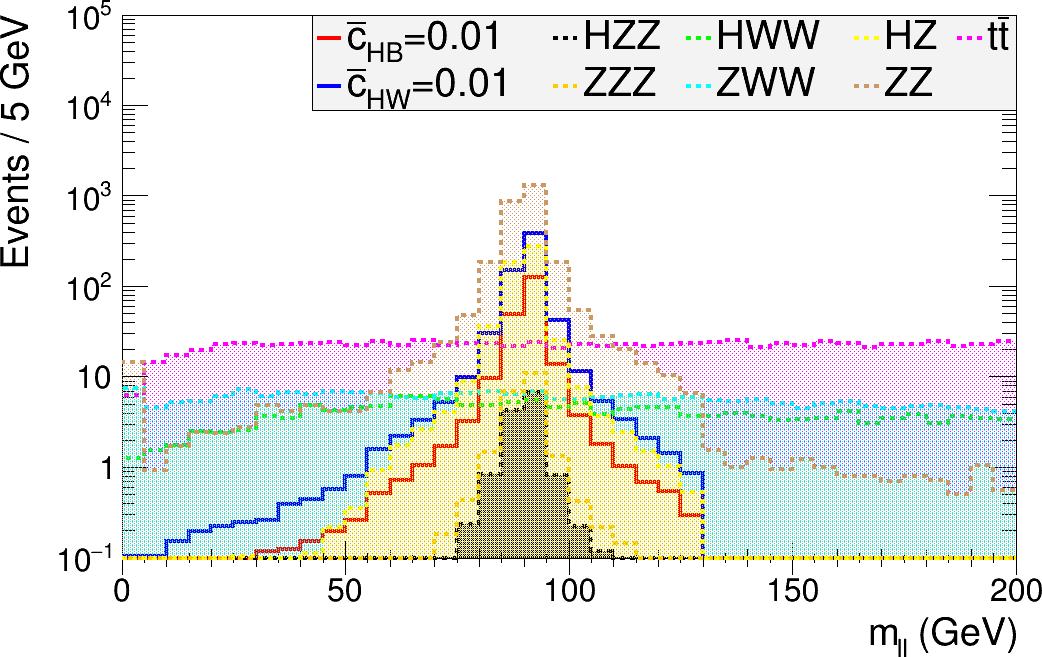}
\caption{}
\label{fig5:b}
\end{subfigure}\hfill

\caption{The invariant mass distributions of two b-tagged jets (a) and charged lepton pair (b) for signal and relevant background processes at CLIC.}
\label{fig5}
\end{figure}

The following kinematic cuts and requirements are applied to separate the signal and background at CLIC using cut-based analysis. As a pre-selection, events containing jet clustering using the VLC algorithm with $N_{j}=2$, $R=1.0$, $\gamma=\beta=1$ and containing $N_{\ell} \geq 2$ same-flavour opposite-sign charged leptons are selected for further analysis (Cut-0). In the exclusive mode, two jets originating from the Higgs boson decay are identified and labeled as b-tagged jets. This selection is denoted as Cut-1. In all kinematic analyzes presented in Figs.~\ref{fig2}-\ref{fig9}, the signal for coefficients $\overline{c}_{HB}$ and $\overline{c}_{HW}$ is generated using the one-parameter analysis and the b-tagging efficiency working point specified in the CLIC Delphes card is applied as the loose WP (90\% b-tagging efficiency). The b-tagged jets with the highest transverse momentum ($p_T$) are labeled $b_1$ (leading), while those with the lower $p_T$ are labeled $b_2$ (sub-leading). Likewise, the charged leptons are ordered by their transverse momentum, with the leading charged lepton ($\ell_1$) having a higher $p_T$ than the sub-leading charged lepton ($\ell_2$). Fig.~\ref{fig2} shows the distributions of the transverse momentum of the leading b-tagged jet, the sub-leading b-tagged jet, the leading charged lepton, and the sub-leading charged lepton for the signal and the relevant background processes. It is evident that the signal can be distinguished from the background by $p_T^{b_1} > 60$ GeV, $p_T^{b_2} > 20$ GeV, $p_T^{\ell_1} > 30$ GeV, $p_T^{\ell_2} > 10$ GeV. Additionally, we consider the pseudo-rapidity of the leading and sub-leading b-tagged jets ($\eta^{b_{1,2}}$), and the leading and sub-leading charged leptons ($\eta^{\ell_{1,2}}$) to be $|\eta^{b_{1,2}}| < 2.5$ and $|\eta^{\ell_{1,2}}| < 2.5$, respectively, and these are labeled Cut-2.

The distributions of the missing energy transverse ($E_T^{miss}$), the transverse momentum balance ratio ($|E_T^{miss}-p_T^{\ell\ell}|/p_T^{\ell\ell}$), the scalar sum of the transverse energy ($H_T$), and the transverse momentum of two b-tagged jets ($p_T^{bb}$) for signal and background processes are shown in Fig.~\ref{fig3}. These figures allow for the determination of the kinematic cuts: $E_T^{miss} > 50$ GeV and $|E_T^{miss}-p_T^{\ell\ell}|/p_T^{\ell\ell} > 0.5$ are labeled Cut-3 and $H_T > 200$ GeV and $p_T^{bb} > 120$ GeV are labeled Cut-4.

In the $\eta – \phi$ space, the angular distance of two particles measured in the detector is defined as $\Delta R=\sqrt{|\Delta\eta|^2+|\Delta\phi|^2}$, where $\Delta\eta$ and $\Delta\phi$ are differences in pseudo-rapidity and azimuthal angle, respectively. The distributions of the angular distance between the two b-tagged jets, between the two charged leptons, and between the leading charged lepton and the leading b-tagged jet are shown in Fig.~\ref{fig4}. For the b-tagged jet and the charged lepton to be properly separated in phase space and identified as distinct objects in the detector, kinematic cuts such as $\Delta R_{bb} < 1.0$, $\Delta R_{\ell \ell} < 3.2$ and $\Delta R_{\ell b} > 0.4$ in Cut-5 are required.

Guided by the invariant mass distributions of the two b-tagged jets and the charged lepton pair in Fig.~\ref{fig5}, we set Cut-6 as $110$ GeV$ < m_{bb} < 140$ GeV and $|m_{\ell\ell}-m_Z| < 4$ GeV, where $m_Z=91.2$ GeV. The b-tagging channel $H \rightarrow b \bar{b}$, which is part of the signal process, is the dominant Higgs boson decay at the LHC with a branching fraction of 53\% \cite{Navas:2024uza}. The invariant mass of the two b-jets produced in this decay is close to 125 GeV, which is the approximate mass of the Higgs boson. As shown in Fig.~\ref{fig5:a}, the $m_{bb}$ invariant mass distribution exhibits a clear peak around 125 GeV, which is consistent with Higgs boson production in the signal process and is essential for its experimental identification of the Higgs boson. A defined cut on the $m_{bb}$ invariant mass will suppress the background processes $e^- e^+ \rightarrow ZZZ$, $e^- e^+ \rightarrow ZZ$, and $e^- e^+ \rightarrow ZWW$, which have a decay of $Z \rightarrow b\bar{b}$. Since the $Z$-boson decays into two charged leptons in the signal process, a large number of events are observed around $m_{Z}\approx 91.2$ GeV according to the $m_{\ell\ell}$ invariant mass distribution in Fig.~\ref{fig5:b}. This resonance peak is a characteristic signature of an on-shell $Z$-boson produced in the experiment and subsequently decaying into a charged lepton pair. The determination of the $m_{\ell\ell}$ invariant mass cut around $m_{Z}$ suppresses background processes that do not involve $Z \rightarrow \ell^-\ell^+$ decay, such as $e^- e^+ \rightarrow HWW$, $e^- e^+ \rightarrow ZWW$, and $e^- e^+ \rightarrow t\bar{t}$. Table~\ref{tab1} lists all the kinematic cuts used in the analysis.

After each cut, the cumulative event numbers of the signal with setting the $\overline{c}_{HB}$ or $\overline{c}_{HW}$ coefficients to $0.001$, respectively, and all other coefficients to zero, and of relevant backgrounds ($HZZ$, $ZZZ$, $HWW$, $ZWW$, $HZ$, $ZZ$, and $t\bar{t}$) are given in Table~\ref{tab2}.  In this table, the integrated luminosity of CLIC are taken into account as ${\cal L}_{\text{int}}=5$ ab$^{-1}$. In Tables~\ref{tab2}-~\ref{tab3}, the expected number of events $N_{exp}$ are calculated using the formula $N_{exp} = \sigma \times \cal L_{\text{int}} \times \epsilon$. Cut efficiency ($\epsilon$) is precisely defined as the number of raw unweighted events surviving after a cut divided by the total number of events produced ($N_{raw}/N_{total}$), where $N_{total}=1$M. The signal-total background ratio ($S/B_{tot}$) where $B_{tot}=B_{HZZ}+B_{ZZZ}+B_{HWW}+B_{ZWW}+B_{HZ}+B_{ZZ}+B_{tt}$ increases by 868 and 1138 times for the $\overline{c}_{HB}$ coefficient and by 1138 and 1237 times for the $\overline{c}_{HW}$ coefficient, respectively, according to the two b-tagging efficiencies of 70\% and 90\% depending on the cuts applied between Cut-0 and Cut-6, thus revealing that the applied cuts play a significant role in suppressing the relevant backgrounds.

\begin{table}[H]
\centering
\caption{Applied kinematic cuts used for the analysis at CLIC and Muon Collider.}
\label{tab1}
\begin{tabular}{p{2.5cm}p{7cm}p{7cm}}
\hline \hline
Cut flow & CLIC at $\sqrt{s}=3$ TeV & Muon Collider at $\sqrt{s}=10$ TeV\\
\hline
Cut-0 & \multicolumn{2}{c}{Jet clustering: Valencia (VLC) algorithm with $R=1.0$,} \\
& \multicolumn{2}{c}{$N_{j}=2$ (exclusive clustering), $N_{\ell} \geq 2$ (least two same flavor opposite-charge leptons)}\\ \hline
Cut-1 & \multicolumn{2}{c}{two b-tagged jets}\\ \hline
Cut-2 & $p_T^{b_1} > 60$ GeV, $p_T^{b_2} > 20$ GeV, & $p_T^{b_1} > 20$ GeV, $p_T^{b_2} > 10$ GeV, \\
& $p_T^{\ell_1} > 30$ GeV, $p_T^{\ell_2} > 10$ GeV, & $p_T^{\ell_1} > 20$ GeV, $p_T^{\ell_2} > 10$ GeV, \\
& $|\eta^{b_{1,2}}| < 2.5$, $|\eta^{\ell_{1,2}}| < 2.5$ & $|\eta^{b_{1,2}}| < 2.5$, $|\eta^{\ell_{1,2}}| < 2.5$\\ \hline
Cut-3 & $E_T^{miss} > 50$ GeV, & $E_T^{miss} > 10$ GeV, \\
& $p_T$ balance: $|E_T^{miss}-p_T^{\ell\ell}|/p_T^{\ell\ell} > 0.5$ & $|E_T^{miss}-p_T^{\ell\ell}|/p_T^{\ell\ell} > 0.5$ \\ \hline
Cut-4 & $H_T > 200$ GeV, $p_T^{bb} > 120$ GeV & $H_T > 150$ GeV, $p_T^{bb} > 140$ GeV\\ \hline
Cut-5 & $\Delta R_{bb} < 1.0$, $\Delta R_{\ell \ell} < 3.2$, $\Delta R_{\ell b} > 0.4$ & $\Delta R_{bb} < 3.0$, $\Delta R_{\ell \ell} < 3.2$, $\Delta R_{\ell b} > 0.4$ \\ \hline
Cut-6 & $110$ GeV$ < m_{bb} < 140$ GeV, & $110$ GeV$ < m_{bb} < 140$ GeV, \\
& $|m_{\ell\ell}-m_Z| < 4$ GeV & $|m_{\ell\ell}-m_Z| < 5$ GeV\\  \hline \hline
\end{tabular}
\end{table}

\begin{table}[H]
\centering
\caption{The cumulative number of events for signal and relevant background processes at CLIC after kinematic cuts.}
\label{tab2}
\begin{tabular}{p{1.0cm}cp{1.0cm}p{1.3cm}p{1.4cm}p{0.1cm}p{1.4cm}p{1.4cm}p{1.4cm}p{1.4cm}p{1.4cm}p{1.4cm}p{1.3cm}}
\hline \hline
& & & \multicolumn{2}{l}{Signals} && \multicolumn{7}{l}{Backgrounds} \\ \cline{4-5} \cline{7-13}
Cuts & b-tag. & Event & $\overline{c}_{HB}$ & $\overline{c}_{HW}$ && $B_{HZZ}$ & $B_{ZZZ}$ & $B_{HWW}$ & $B_{ZWW}$ & $B_{HZ}$ & $B_{ZZ}$ & $B_{tt}$ \\ \hline
\multirow{2}{*}{Cut-0} & \multirow{2}{*}{-} & $N_{exp}$ & 9.39361 & 14.8028 && 8.17458 & 16.0476 & 1013.50 & 454.209 & 1.14242 & 1666.63 & 1235.21\\ [-6pt]
& & $N_{raw}$ & 571528 & 517724 && 596172 & 696814 & 395621 & 396164 & 2100 & 571523 & 303007\\ \hline

\multirow{4}{*}{Cut-1} & \multirow{2}{*}{70\%} & $N_{exp}$ & 3.70927 & 6.21946 && 3.14235 & 4.81824 & 438.488 & 160.347 & 0.31334 & 186.340 & 379.050\\ [-6pt]
& & $N_{raw}$ & 225680 & 217524 && 229172 & 209216 & 171164 & 139856 & 576 & 63900 & 92984\\ 
& \multirow{2}{*}{90\%} & $N_{exp}$ & 6.26611 & 10.3798 && 5.32988 & 8.23183 & 730.082 & 269.230 & 0.53530 & 339.810 & 651.197\\ [-6pt]
& & $N_{raw}$ & 381244 & 363031 && 388708 & 357440 & 284988 & 234824 & 984 & 116528 & 159744\\  \hline

\multirow{4}{*}{Cut-2} & \multirow{2}{*}{70\%} & $N_{exp}$ & 3.20093 & 5.54903 && 2.67824 & 3.24197 & 369.965 & 90.7354 & 0.28723 & 12.5510 & 316.793\\ [-6pt]
& & $N_{raw}$ & 194752 & 194076 && 195324 & 140772 & 144416 & 79140 & 528 & 4304 & 77712\\ 
& \multirow{2}{*}{90\%} & $N_{exp}$ & 5.35187 & 9.18732 && 4.48847 & 5.43433 & 615.487 & 150.996 & 0.49396 & 20.9027 & 543.317\\ [-6pt]
& & $N_{raw}$ & 325620 & 321324 && 327344 & 235968 & 240256 & 131700 & 908 & 7168 & 133280\\  \hline

\multirow{4}{*}{Cut-3} & \multirow{2}{*}{70\%} & $N_{exp}$ & 2.22562 & 4.24329 && 1.74715 & 1.90172 & 212.342 & 56.6976 & 0.15232 & 5.54063 & 213.870\\ [-6pt]
& & $N_{raw}$ & 135412 & 148408 && 127420 & 82576 & 82888 & 49452 & 280 & 1900 & 52464\\  
& \multirow{2}{*}{90\%} & $N_{exp}$ & 3.72110 & 7.03251 && 2.92653 & 3.19186 & 352.944 & 94.4272 & 0.31334 & 9.28493 & 365.989\\ [-6pt]
& & $N_{raw}$ & 226400 & 245960 && 213432 & 138596 & 137772 & 82360 & 576 & 3184 & 89780\\  \hline

\multirow{4}{*}{Cut-4} & \multirow{2}{*}{70\%} & $N_{exp}$ & 2.11149 & 4.12549 && 1.63115 & 1.68487 & 203.796 & 44.4574 & 0.15232 & 4.73578 & 176.790\\ [-6pt]
& & $N_{raw}$ & 128468 & 144288 && 118960 & 73160 & 79552 & 38776 & 280 & 1624 & 43368\\  
& \multirow{2}{*}{90\%} & $N_{exp}$ & 3.51256 & 6.82939 && 2.72574 & 2.81601 & 338.496 & 73.7990 & 0.31334 & 7.78021 & 303.765\\ [-6pt]
& & $N_{raw}$ & 213712 & 238856 && 198788 & 122276 & 132132 & 64368 & 576 & 2668 & 74516\\  \hline

\multirow{4}{*}{Cut-5} & \multirow{2}{*}{70\%} & $N_{exp}$ & 1.85877 & 3.79165 && 1.38878 & 1.50938 & 138.265 & 28.4978 & 0.13926 & 4.33919 & 3.48949\\ [-6pt]
& & $N_{raw}$ & 113092 & 132612 && 101284 & 65540 & 53972 & 24856 & 256 & 1488 & 856\\ 
& \multirow{2}{*}{90\%} & $N_{exp}$ & 3.08569 & 6.26647 && 2.32146 & 2.51625 & 229.240 & 47.1631 & 0.26982 & 7.13866 & 6.35935\\ [-6pt]
& & $N_{raw}$ & 187740 & 219168 && 169304 & 109260 & 89484 & 41136 & 496 & 2448 & 1560\\  \hline

\multirow{4}{*}{Cut-6} & \multirow{2}{*}{70\%} & $N_{exp}$ & 0.85204 & 1.76287 && 0.06530 & 0.00617 & 0.29716 & 0.04127 & 0.00870 & 0.03499 & 0.00742\\ [-6pt]
& & $N_{raw}$ & 51840 & 61656 && 4762 & 268 & 116 & 36 & 16 & 12 & 2\\  
& \multirow{2}{*}{90\%} & $N_{exp}$ & 1.40278 & 2.90953 && 0.11018 & 0.00958 & 0.45087 & 0.04586 & 0.01523 & 0.05832 & 0.00991\\ [-6pt]
& & $N_{raw}$ & 85348 & 101760 && 8035 & 416 & 176 & 40 & 28 & 20 & 3\\ 
\hline \hline
\end{tabular}
\end{table}

\subsection{Muon Collider Study}

\begin{figure}[H]
\centering
\begin{subfigure}{0.47\linewidth}
\includegraphics[width=\linewidth]{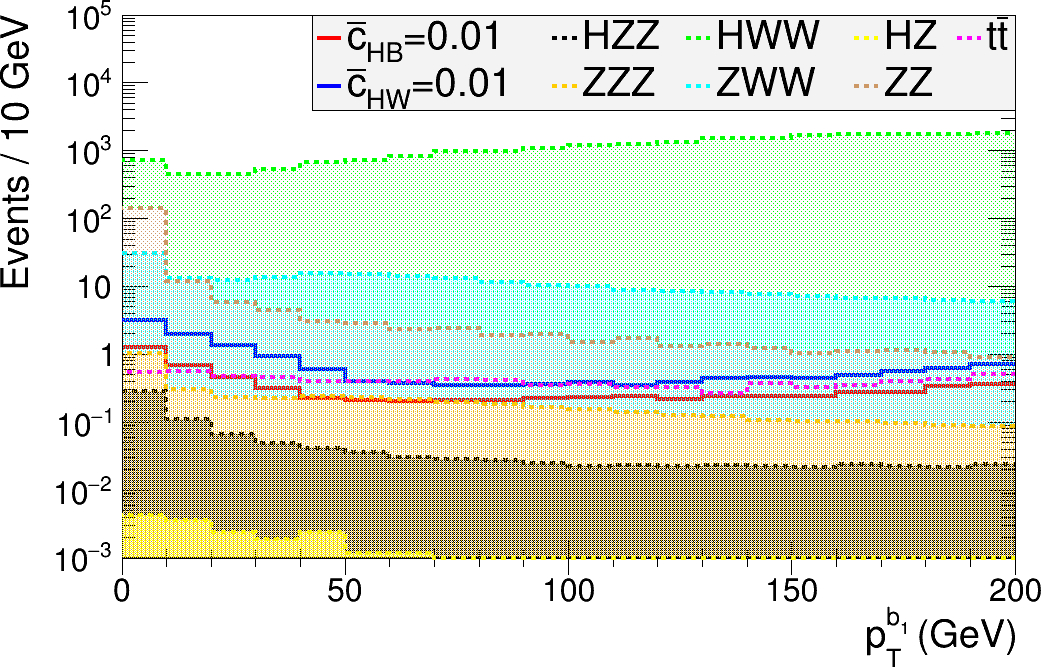}
\caption{}
\label{fig6:a}
\end{subfigure}\hfill
\begin{subfigure}{0.47\linewidth}
\includegraphics[width=\linewidth]{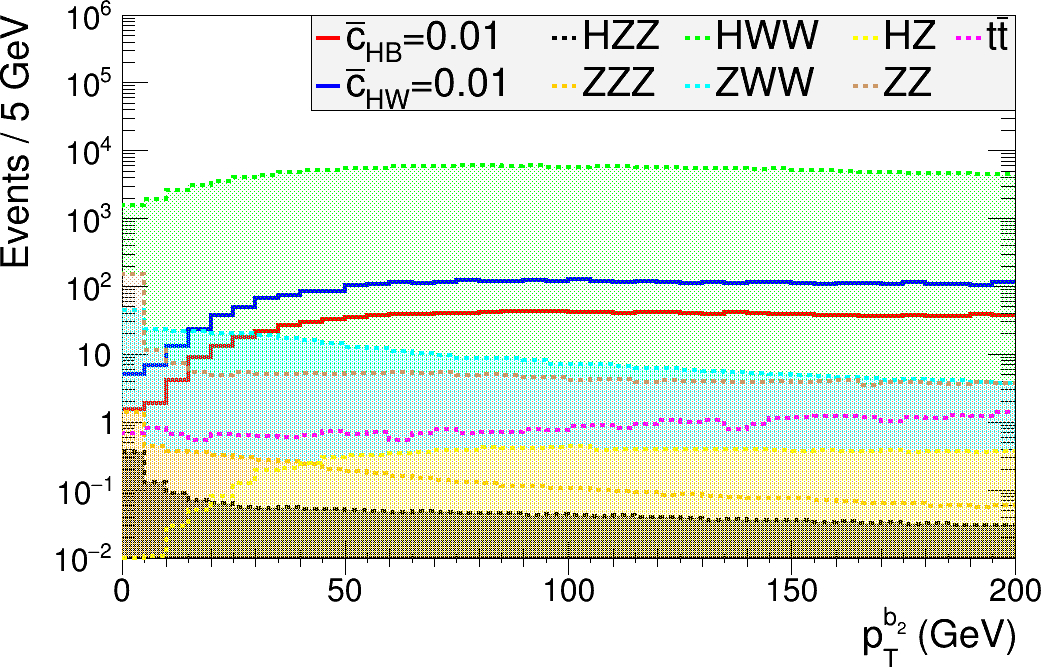}
\caption{}
\label{fig6:b}
\end{subfigure}\hfill

\begin{subfigure}{0.47\linewidth}
\includegraphics[width=\linewidth]{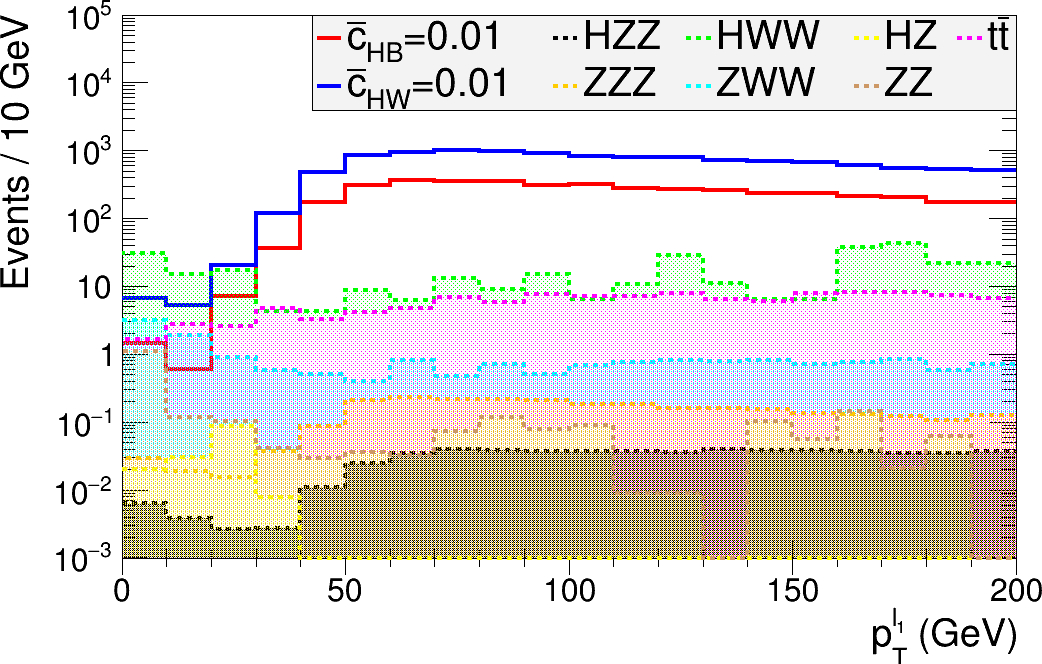}
\caption{}
\label{fig6:c}
\end{subfigure}\hfill
\begin{subfigure}{0.47\linewidth}
\includegraphics[width=\linewidth]{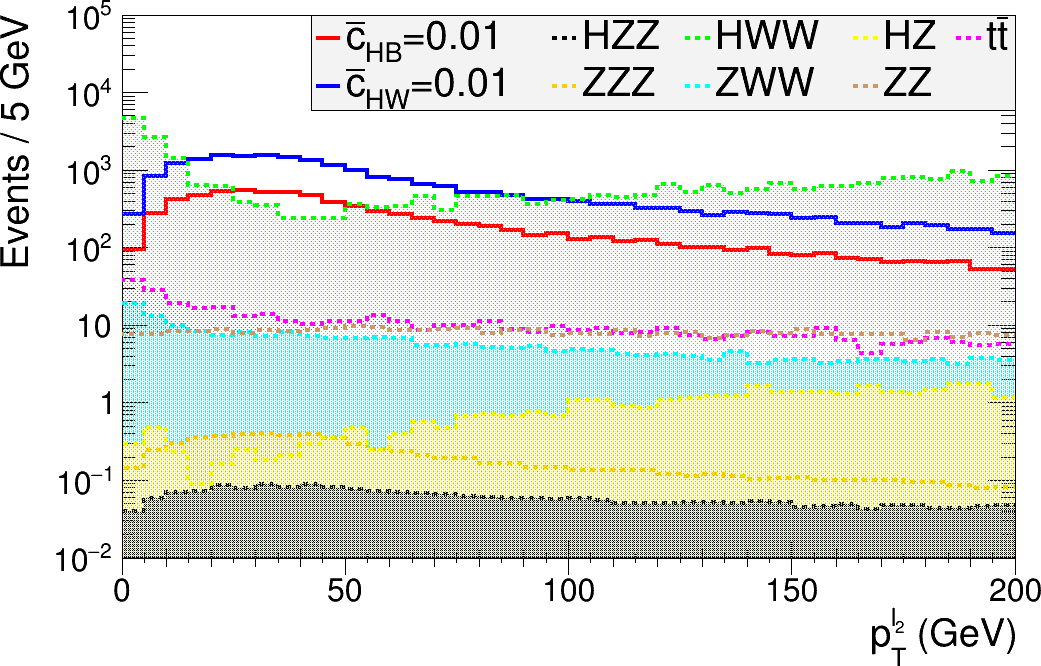}
\caption{}
\label{fig6:d}
\end{subfigure}\hfill

\caption{The distributions of transverse momentum of the leading b-tagged jet (a), the sub-leading b-tagged jet (b), the leading charged lepton (c) and the sub-leading charged lepton (d) for signal and relevant background
processes at Muon Collider.}
\label{fig6}
\end{figure}

The pre-selection for the Muon Collider study analyses is considered the same as for the CLIC study (Cut-0). In the exclusive mode, the selection of Higgs boson decay to two jets as b-tagged jets is called Cut-1. Fig.~\ref{fig6} presents the transverse momentum distributions of the leading b-tagged jet, the sub-leading b-tagged jet, the leading charged lepton, and the sub-leading charged lepton for the signal process $\mu^- \mu^+ \rightarrow HZZ$ and the relevant background processes at Muon Collider. The signal can be distinguished from the background with $p_T^{b_1} > 20$ GeV, $p_T^{b_2} > 10$ GeV, $p_T^{\ell_1} > 20$ GeV, $p_T^{\ell_2} > 10$ GeV according to this figure and furthermore, the pseudo-rapidity of the leading and sub-leading b-tagged jets ($\eta^{b_{1,2}}$), and the pseudo-rapidity of the leading and sub-leading charged leptons ($\eta^{\ell_{1,2}}$) are $|\eta^{b_{1,2}}| < 2.5$ and $|\eta^{\ell_{1,2}}| < 2.5$, respectively, which are labeled as Cut-2.

\begin{figure}[H]
\centering
\begin{subfigure}{0.48\linewidth}
\includegraphics[width=\linewidth]{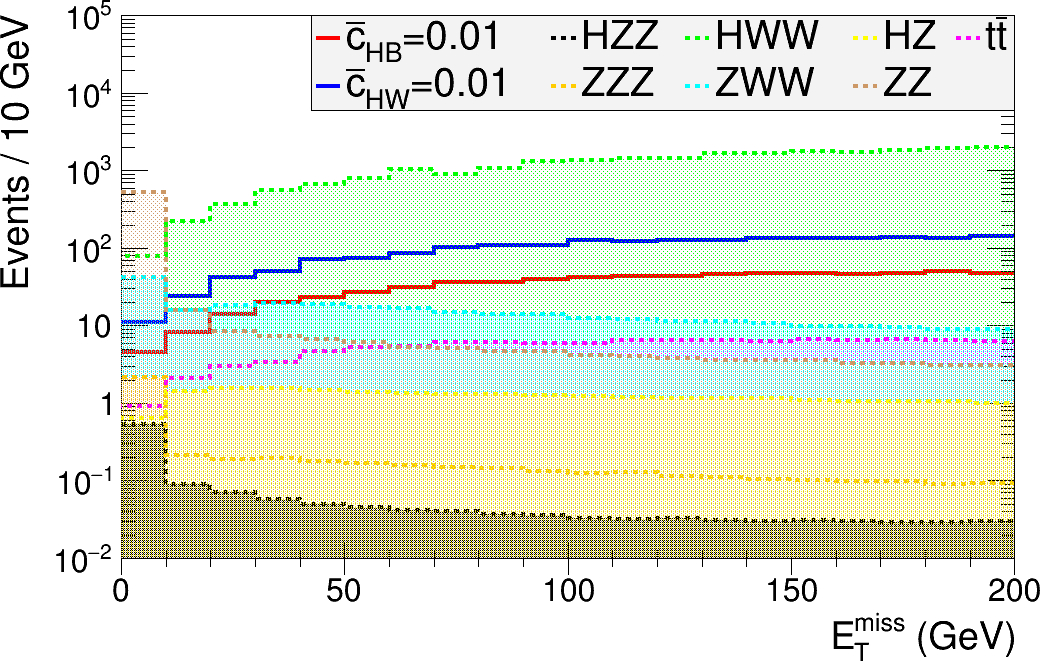}
\caption{}
\label{fig7:a}
\end{subfigure}\hfill
\begin{subfigure}{0.48\linewidth}
\includegraphics[width=\linewidth]{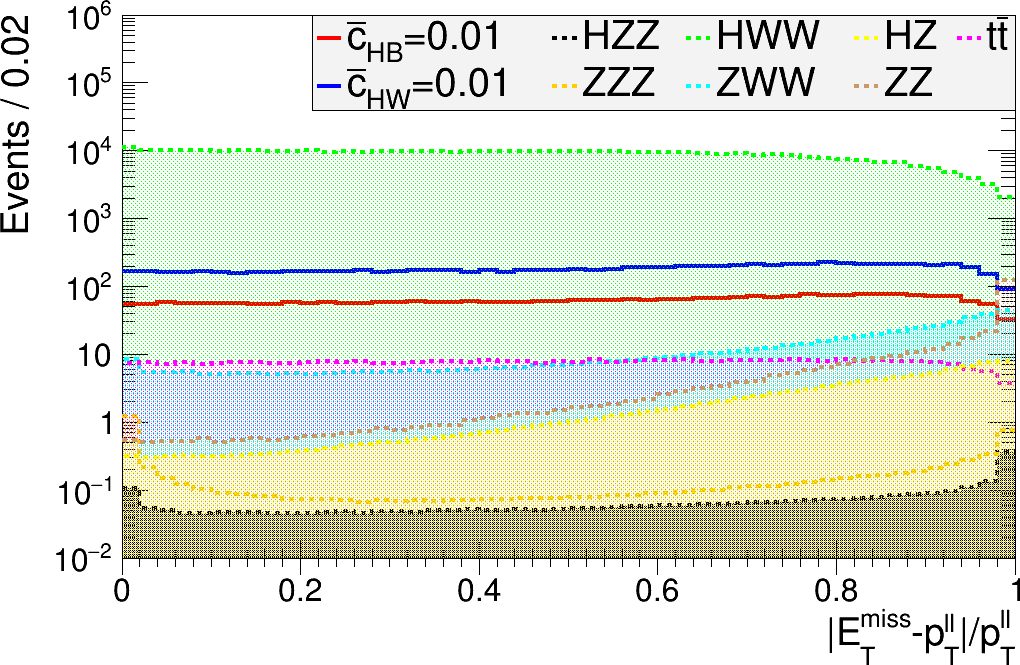}
\caption{}
\label{fig7:b}
\end{subfigure}\hfill

\begin{subfigure}{0.48\linewidth}
\includegraphics[width=\linewidth]{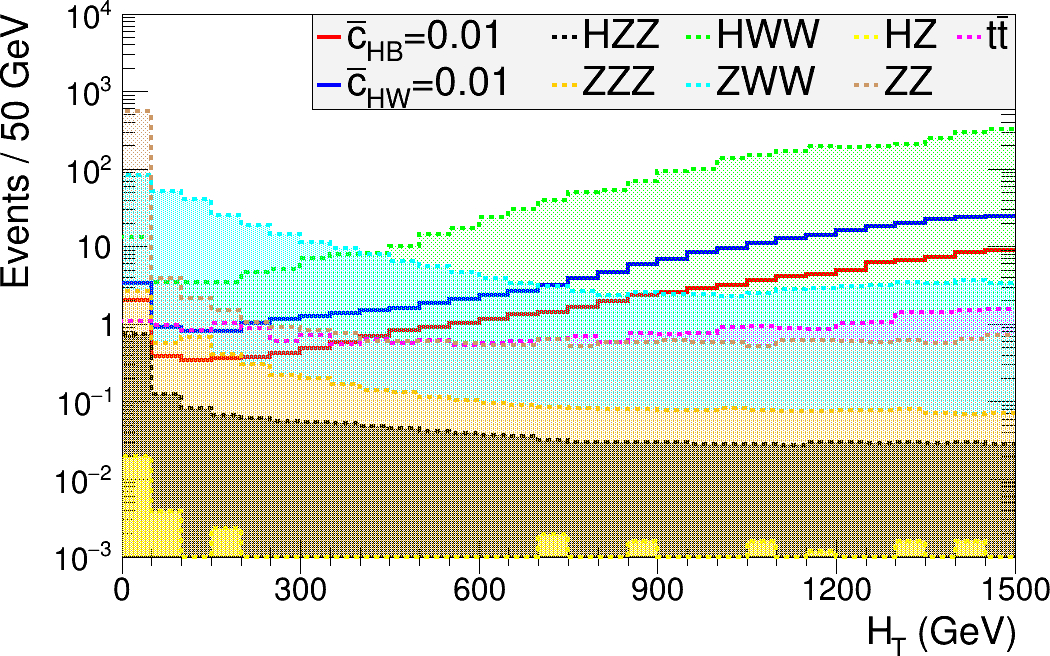}
\caption{}
\label{fig7:c}
\end{subfigure}\hfill
\begin{subfigure}{0.48\linewidth}
\includegraphics[width=\linewidth]{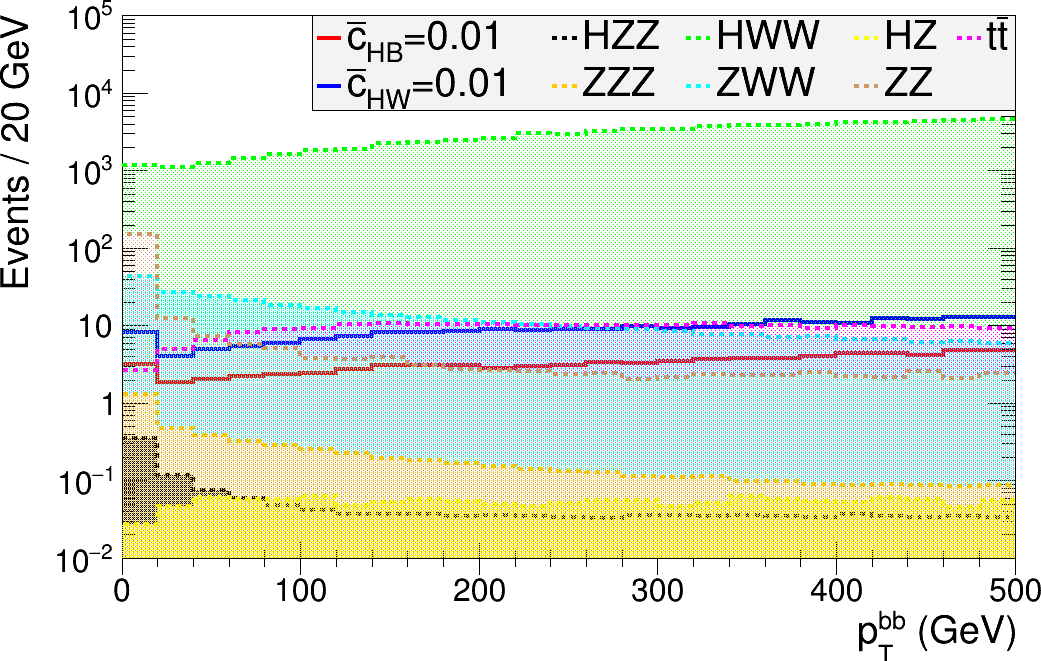}
\caption{}
\label{fig7:d}
\end{subfigure}\hfill

\caption{The distributions of the missing energy transverse (a), the transverse momentum balance ratio (b), the scalar sum of the transverse energy (c), and the transverse momentum of two b-tagged jets (d) for signal and relevant background processes at Muon Collider.}
\label{fig7}
\end{figure}

The distributions of the missing energy transverse ($E_T^{miss}$), the transverse momentum balance ratio ($|E_T^{miss}-p_T^{\ell\ell}|/p_T^{\ell\ell}$), the scalar sum of the transverse energy ($H_T$), and the transverse momentum of two b-tagged jets ($p_T^{bb}$) are given in Fig.~\ref{fig7}. We determine the kinematic cuts based on these figures: $E_T^{miss} > 10$ GeV and $|E_T^{miss}-p_T^{\ell\ell}|/p_T^{\ell\ell} > 0.5$ are labeled Cut-3 and $H_T > 150$ GeV and $p_T^{bb} > 140$ GeV are labeled Cut-4.

\begin{figure}[H]
\centering
\begin{subfigure}{0.48\linewidth}
\includegraphics[width=\linewidth]{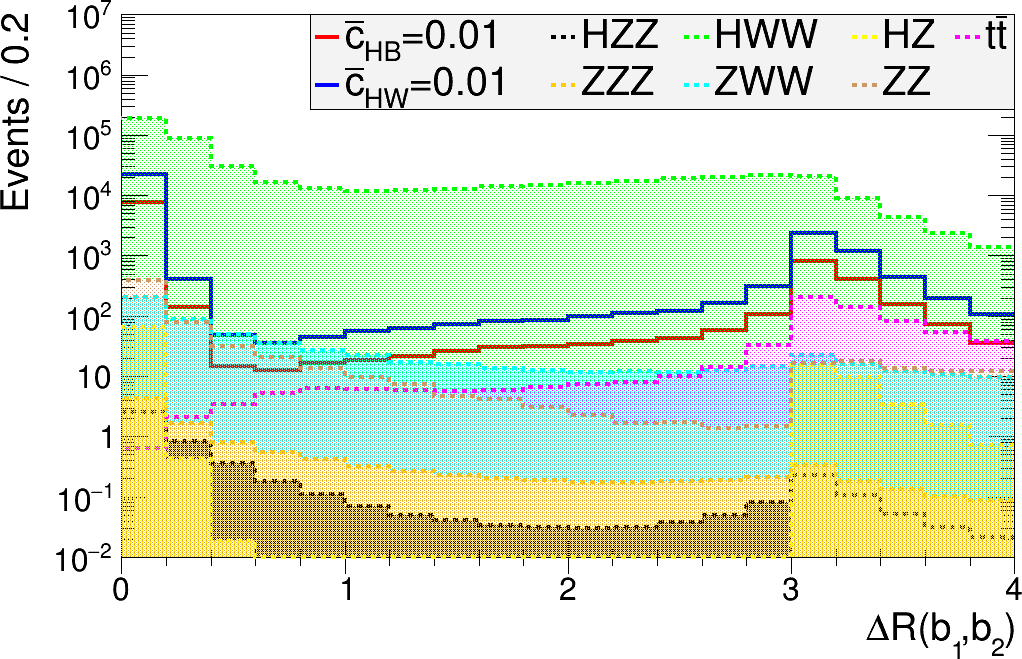}
\caption{}
\label{fig8:a}
\end{subfigure}\hfill
\begin{subfigure}{0.48\linewidth}
\includegraphics[width=\linewidth]{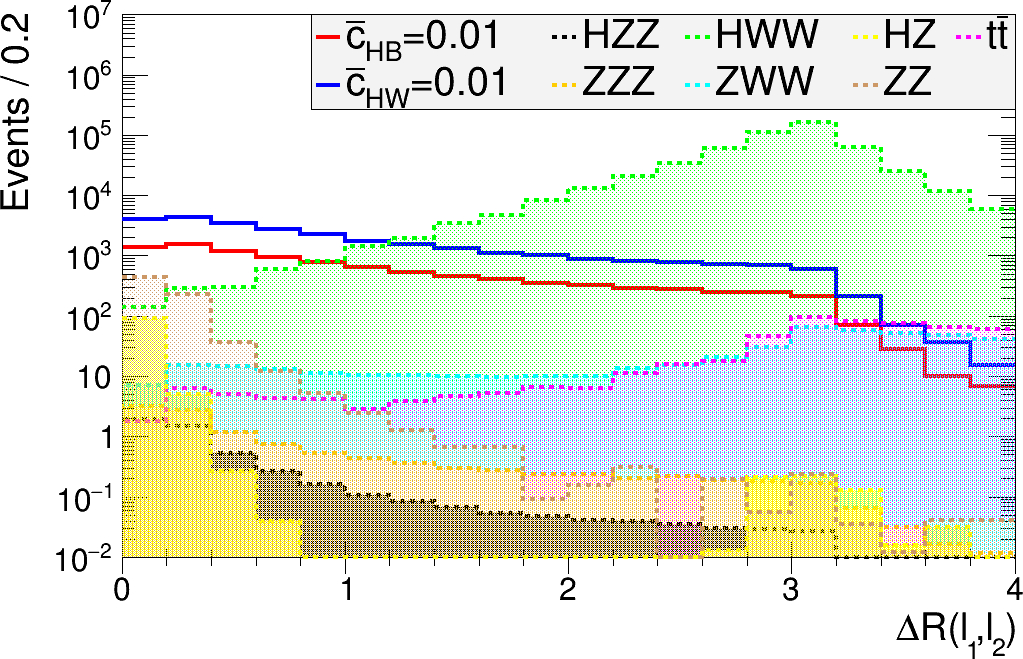}
\caption{}
\label{fig8:b}
\end{subfigure}\hfill
\begin{subfigure}{0.48\linewidth}
\includegraphics[width=\linewidth]{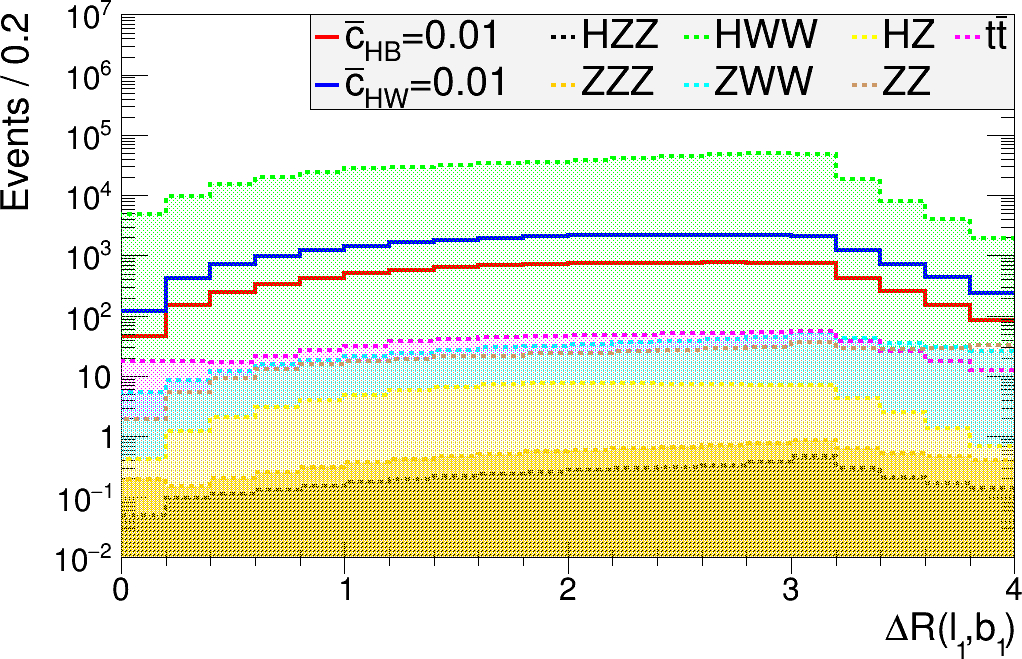}
\caption{}
\label{fig8:c}
\end{subfigure}\hfill

\caption{The distributions of the angular distance between the two b-tagged jets (a), between the two charged leptons (b), and between the leading charged lepton and the leading b-tagged jet (c) for signal and relevant background processes at Muon Collider.}
\label{fig8}
\end{figure}

Fig.~\ref{fig8} shows the distributions of the angular distance between the two b-tagged jets, between the two charged leptons, and between the leading charged lepton and the leading b-tagged jet at Muon Collider. According to these figures, kinematic cuts of $\Delta R_{bb} < 3.0$, $\Delta R_{\ell \ell} < 3.2$ and $\Delta R_{\ell b} > 0.4$ are applied in Cut-5.

\begin{figure}[H]
\centering
\begin{subfigure}{0.48\linewidth}
\includegraphics[width=\linewidth]{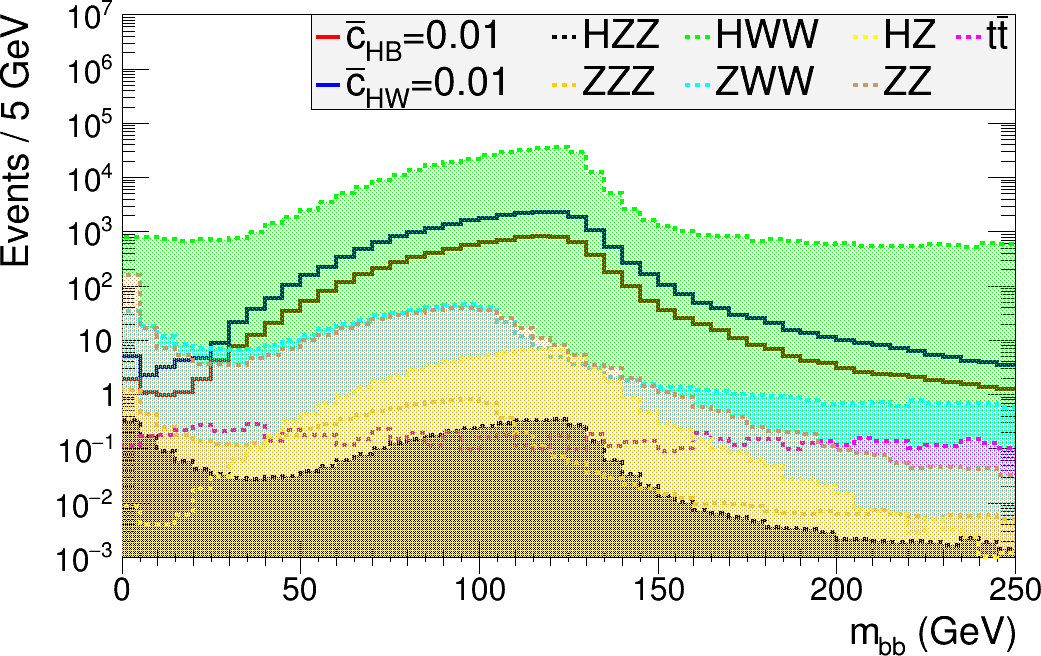}
\caption{}
\label{fig9:a}
\end{subfigure}\hfill
\begin{subfigure}{0.48\linewidth}
\includegraphics[width=\linewidth]{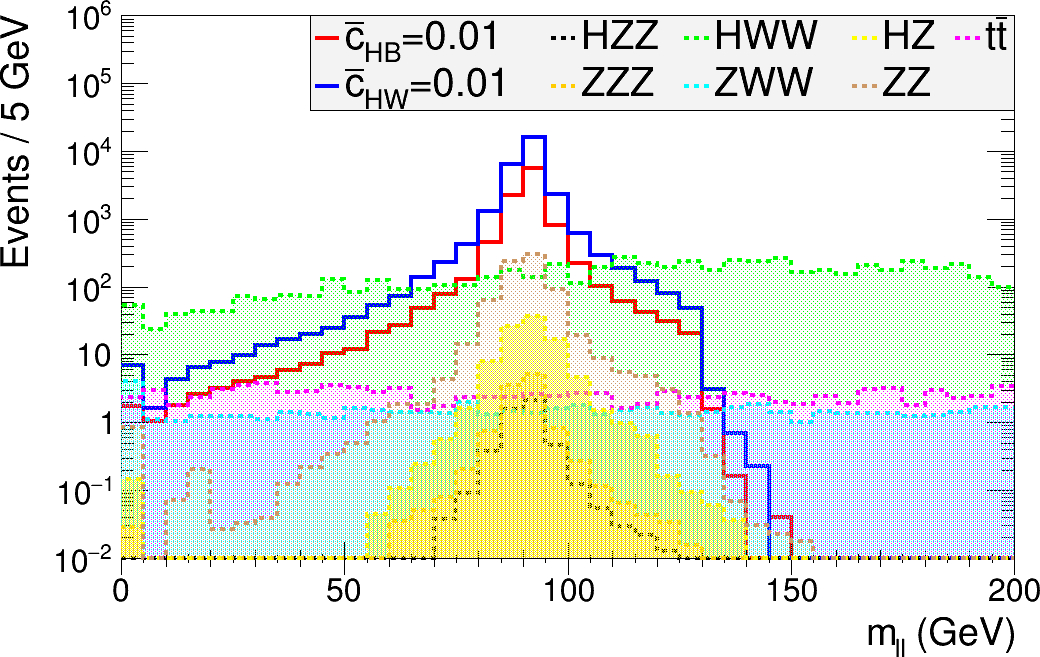}
\caption{}
\label{fig9:b}
\end{subfigure}\hfill

\caption{The invariant mass distributions of two b-tagged jets (a) and charged lepton pair (b) for signal and relevant background processes at Muon Collider.}
\label{fig9}
\end{figure}

Using the invariant mass distributions of the two b-labeled jets and the charged lepton pair in Fig.~\ref{fig9}, we defined Cut-6 as $110$ GeV$ < m_{bb} < 140$ GeV and $|m_{\ell\ell}-m_Z| < 5$ GeV, where $m_Z=91.2$ GeV to separate the signal from the backgrounds. Table~\ref{tab1} summarizes all kinematic cuts used in the analysis for the Muon Collider.

Table~\ref{tab3} lists the cumulative event numbers after each cut for the evaluated signal and the relevant background processes ($HZZ$, $ZZZ$, $HWW$, $ZWW$, $HZ$, $ZZ$, and $t\bar{t}$), where $\overline{c}_{HB}$ or $\overline{c}_{HW}$ coefficient is set to $0.0002$ while all other parameters are fixed to zero. The results are presented assuming an integrated luminosity of ${\cal L}_{\text{int}}=10$ ab$^{-1}$ for Muon Collider. As kinematic cuts from Cut-0 to Cut-6 are applied, the signal-total-background ratio ($S/B_{tot}$), where $B_{tot}=B_{HZZ}+B_{ZZZ}+B_{HWW}+B_{ZWW}+B_{HZ}+B_{ZZ}+B_{tt}$, improves by factors of 8561 and 7513 for $\overline{c}_{HB}$ coefficient, and by 8313 and 7270 for $\overline{c}_{HW}$ coefficient, respectively, according to the two b-tagging efficiencies of 70\% and 90\%. These results clearly demonstrate the effectiveness of the applied cuts in reducing the background contributions.

\begin{table}[H]
\centering
\caption{The cumulative number of events for signal and relevant background processes at Muon Collider after kinematic cuts.}
\label{tab3}
\begin{tabular}{p{1.0cm}cp{1.0cm}p{1.3cm}p{1.4cm}p{0.1cm}p{1.4cm}p{1.4cm}p{1.4cm}p{1.4cm}p{1.4cm}p{1.4cm}p{1.3cm}}
\hline \hline
& & & \multicolumn{2}{l}{Signals} && \multicolumn{7}{l}{Backgrounds} \\ \cline{4-5} \cline{7-13}
Cuts & b-tag. & Event & $\overline{c}_{HB}$ & $\overline{c}_{HW}$ && $B_{HZZ}$ & $B_{ZZZ}$ & $B_{HWW}$ & $B_{ZWW}$ & $B_{HZ}$ & $B_{ZZ}$ & $B_{tt}$ \\ \hline
\multirow{2}{*}{Cut-0} & \multirow{2}{*}{-} & $N_{exp}$ & 7.91626 & 20.1753 && 4.03671 & 9.62071 & 230222 & 294.294 & 23.3205 & 287.107 & 307.096\\ [-6pt]
& & $N_{raw}$ & 843124 & 861372 && 811061 & 848956 & 423887 & 439208 & 238908 & 386715 & 418430\\ \hline

\multirow{4}{*}{Cut-1} & \multirow{2}{*}{70\%} & $N_{exp}$ & 3.17757 & 8.20323 && 1.55928 & 3.38639 & 77490.3 & 101.916 & 9.45244 & 64.5226 & 107.953\\ [-6pt]
& & $N_{raw}$ & 338428 & 350232 && 313292 & 298824 & 142676 & 152101 & 96836 & 86908 & 147090\\ 
& \multirow{2}{*}{90\%} & $N_{exp}$ & 5.31996 & 13.6987 && 2.61916 & 5.69213 & 129580 & 173.143 & 15.6762 & 114.693 & 185.411\\ [-6pt]
& & $N_{raw}$ & 566604 & 584858 && 526245 & 502288 & 238584 & 258400 & 160596 & 154484 & 252630\\  \hline

\multirow{4}{*}{Cut-2} & \multirow{2}{*}{70\%} & $N_{exp}$ & 3.00853 & 7.84047 && 1.45418 & 3.04134 & 75215.7 & 86.0487 & 9.31695 & 48.0646 & 93.7661\\ [-6pt]
& & $N_{raw}$ & 320424 & 334744 && 292176 & 268376 & 138488 & 128420 & 95448 & 64740 & 127760\\ 
& \multirow{2}{*}{90\%} & $N_{exp}$ & 5.01800 & 13.0816 && 2.42001 & 5.05670 & 125587 & 145.354 & 15.4787 & 80.7906 & 160.861\\ [-6pt]
& & $N_{raw}$ & 534444 & 558511 && 486231 & 446216 & 231232 & 216928 & 158572 & 108820 & 219179\\  \hline

\multirow{4}{*}{Cut-3} & \multirow{2}{*}{70\%} & $N_{exp}$ & 2.34411 & 6.53753 && 0.98912 & 1.68753 & 42107.1 & 66.3115 & 7.70048 & 13.9605 & 75.0291\\ [-6pt]
& & $N_{raw}$ & 249660 & 279116 && 198735 & 148912 & 77528 & 98964 & 78888 & 18804 & 102230\\  
& \multirow{2}{*}{90\%} & $N_{exp}$ & 3.91011 & 10.9016 && 1.64944 & 2.80323 & 70523.1 & 112.168 & 12.7826 & 23.2884 & 128.150\\ [-6pt]
& & $N_{raw}$ & 416447 & 465437 && 331408 & 247364 & 129848 & 167401 & 130952 & 31368 & 174609\\  \hline

\multirow{4}{*}{Cut-4} & \multirow{2}{*}{70\%} & $N_{exp}$ & 2.26986 & 6.46370 && 0.90403 & 1.28935 & 41640.0 & 41.9616 & 7.69892 & 11.0354 & 69.7008\\ [-6pt]
& & $N_{raw}$ & 241752 & 275964 && 181639 & 113776 & 76668 & 62624 & 78872 & 14864 & 94970\\  
& \multirow{2}{*}{90\%} & $N_{exp}$ & 3.78227 & 10.7739 && 1.50509 & 2.14250 & 69682.4 & 71.0099 & 12.7795 & 18.5220 & 119.094\\ [-6pt]
& & $N_{raw}$ & 402832 & 459985 && 302405 & 189060 & 128300 & 105976 & 130920 & 24948 & 162270\\  \hline

\multirow{4}{*}{Cut-5} & \multirow{2}{*}{70\%} & $N_{exp}$ & 2.19362 & 6.26274 && 0.86358 & 1.24647 & 32111.5 & 25.5667 & 7.17884 & 10.6375 & 1.65133\\ [-6pt]
& & $N_{raw}$ & 233632 & 267384 && 173512 & 109992 & 59124 & 38156 & 73544 & 14328 & 2250\\ 
& \multirow{2}{*}{90\%} & $N_{exp}$ & 3.64342 & 10.3901 && 1.43310 & 2.05978 & 53471.3 & 42.9265 & 11.8377 & 17.4915 & 3.79439\\ [-6pt]
& & $N_{raw}$ & 388044 & 443599 && 287940 & 181760 & 98452 & 64064 & 121272 & 23560 & 5170\\  \hline

\multirow{4}{*}{Cut-6} & \multirow{2}{*}{70\%} & $N_{exp}$ & 0.74884 & 2.18136 && 0.04079 & 0.00815 & 2.17248 & 0.04288 & 0.48064 & 0.16927 & 0.00132\\ [-6pt]
& & $N_{raw}$ & 79755 & 93132 && 8196 & 719 & 4 & 64 & 4924 & 228 & 2\\  
& \multirow{2}{*}{90\%} & $N_{exp}$ & 1.24381 & 3.61078 && 0.06804 & 0.01328 & 4.34496 & 0.05360 & 0.75435 & 0.28212 & 0.00198\\ [-6pt]
& & $N_{raw}$ & 132472 & 154160 && 13671 & 1172 & 8 & 80 & 7728 & 380 & 3\\ 
\hline \hline
\end{tabular}
\end{table}

\section{Sensitivities on the anomalous Higgs-gauge boson couplings} \label{Sec4}

The $\chi^2$ test is used to calculate the sensitivity of dimension-six Higgs-gauge boson couplings in the process $\ell^- \ell^+ \rightarrow HZZ$. The analysis utilizes the $\chi^2$ distribution to determine limits at 95\% Confidence Level (C.L.) with a threshold value of $3.84$ for one degree of freedom. The $\chi^2$ distribution is defined as follows:

\begin{eqnarray}
\label{eq.12}
\chi^{2}=\sum_{i}^{n_{bins}} (\frac{N_{i}^{NP}-N_{i}^{B}}{N_{i}^{B}\Delta_{i}})^{2}
\end{eqnarray}

{\raggedright where by integrating the invariant mass $m_{\ell \ell bb}$ distribution of $\ell\ell bb$ system  after Cut-6, $N_{i}^{B}$ is the number of events of relevant total backgrounds in ith bin and $N_{i}^{NP}$ is the total number of events in ith bin, including both contribution of relevant total backgrounds and signal contributions from effective couplings for a given value of Wilson coefficients $\overline{c}_{HB}$ and $\overline{c}_{HW}$. The $N_{i}^{NP}$ values obtained after applying the cuts have a different dependence on the relevant value of the Wilson coefficients $\overline{c}_{HB}$ and $\overline{c}_{HW}$ than the total cross section.  $\Delta_{i}=\sqrt{1/N_i^B}$ is the statistical uncertainty in each bin. To determine the 95\% C.L. limits, total cross-sections ($\sigma_{tot} = \sigma_{SM} + \sigma_{int} + \sigma_{dim6}$) including the SM cross-section, the linear interference terms between the SM and dimension-six terms and the quadratic dimension-six terms are calculated using MadGraph5 for various values of the coefficients $\overline{c}_{HB}$ and $\overline{c}_{HW}$. A function $\chi^2$ is constructed from these cross-sections using Eq.~(\ref{eq.12}), and the corresponding solution in 3.84 yields the limit values.}

\begin{figure}[H]
\centering
\begin{subfigure}{0.48\linewidth}
\includegraphics[width=\linewidth]{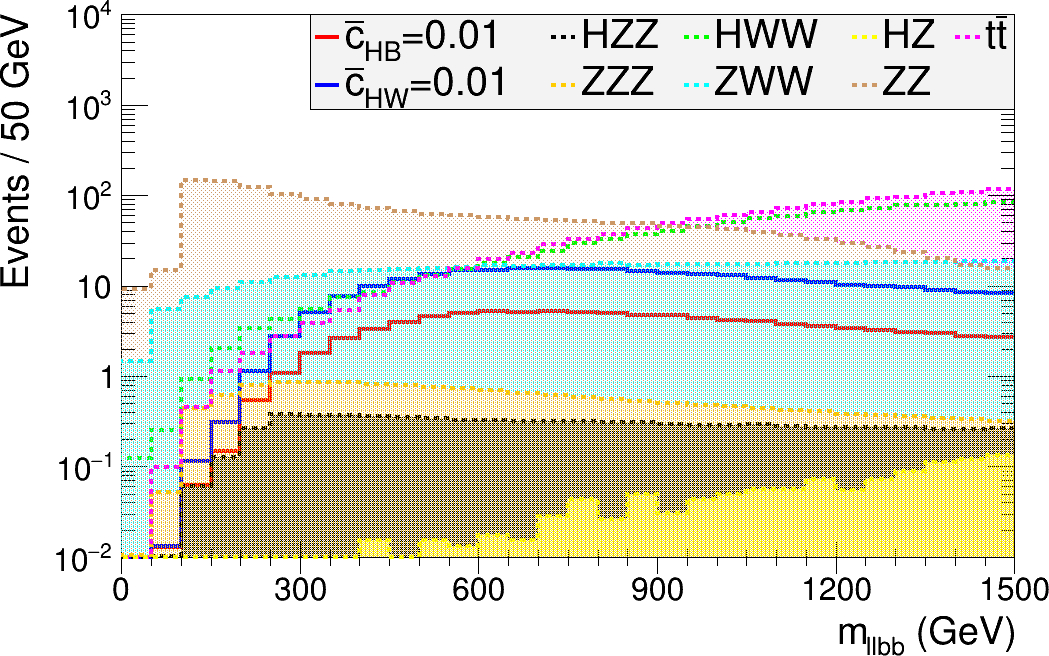}
\caption{}
\label{fig10:a}
\end{subfigure}\hfill
\begin{subfigure}{0.48\linewidth}
\includegraphics[width=\linewidth]{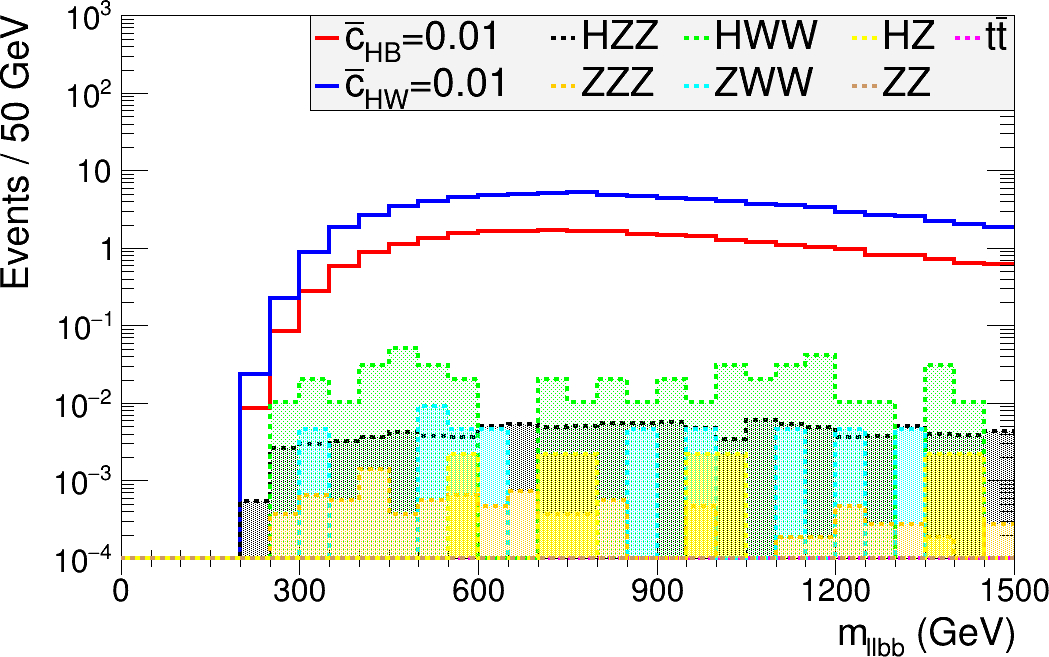}
\caption{}
\label{fig10:b}
\end{subfigure}\hfill

\begin{subfigure}{0.48\linewidth}
\includegraphics[width=\linewidth]{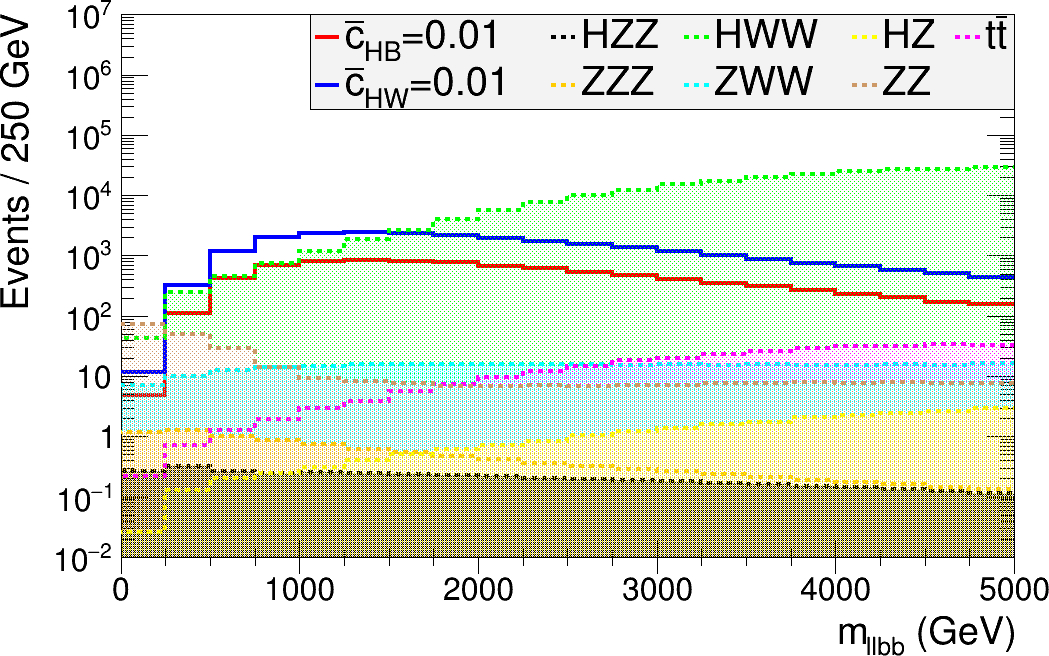}
\caption{}
\label{fig10:c}
\end{subfigure}\hfill
\begin{subfigure}{0.48\linewidth}
\includegraphics[width=\linewidth]{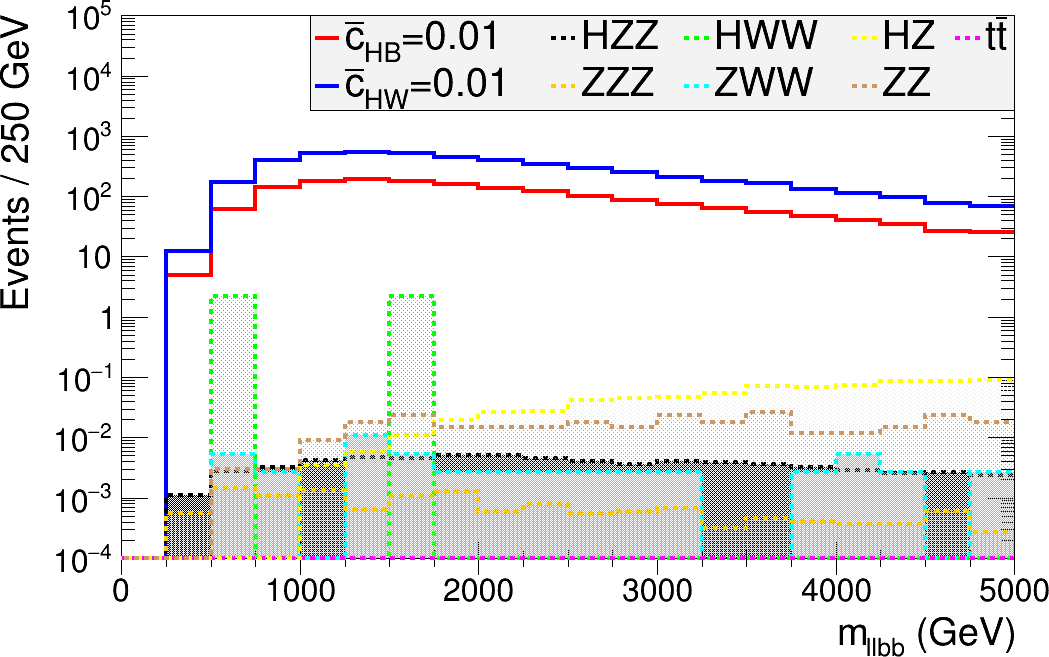}
\caption{}
\label{fig10:d}
\end{subfigure}\hfill

\caption{The invariant mass distributions of the two leptons and two b-jets system after Cut-0 (a,c) and after Cut-6 (b,d) for the signal and relevant background processes at CLIC (a,b) and Muon Collider (c,d).}
\label{fig10}
\end{figure}

Fig.~\ref{fig10} shows the four-body invariant mass distributions of the two leptons and two b-jets system after Cut-0 (left) and after Cut-6 (right) for the signal and relevant background processes at CLIC (top) and Muon Collider (bottom).

\begin{figure}[H]
\centering
\begin{subfigure}{0.485\linewidth}
\includegraphics[width=\linewidth]{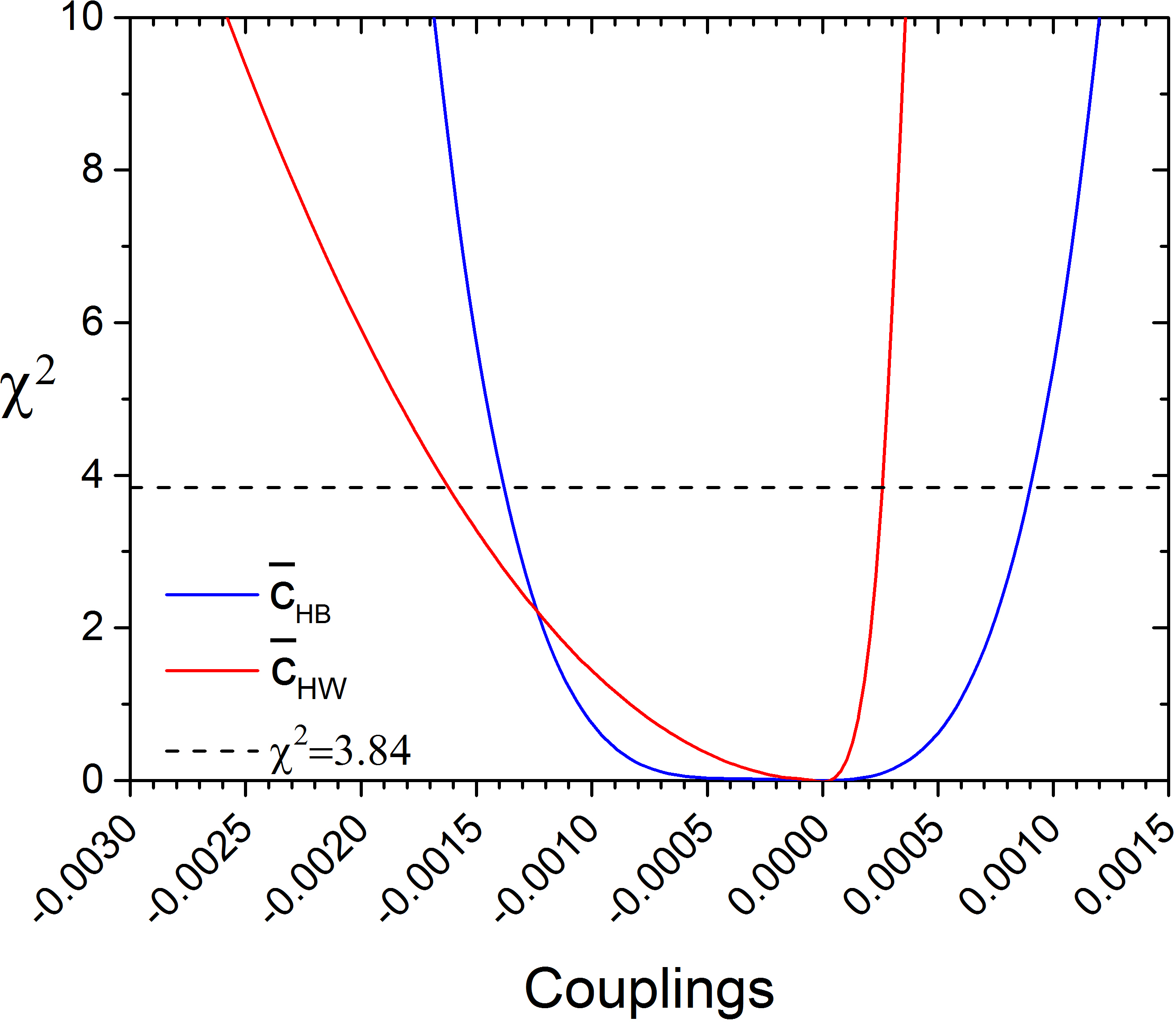}
\caption{}
\label{fig11:a}
\end{subfigure}\hfill
\begin{subfigure}{0.5\linewidth}
\includegraphics[width=\linewidth]{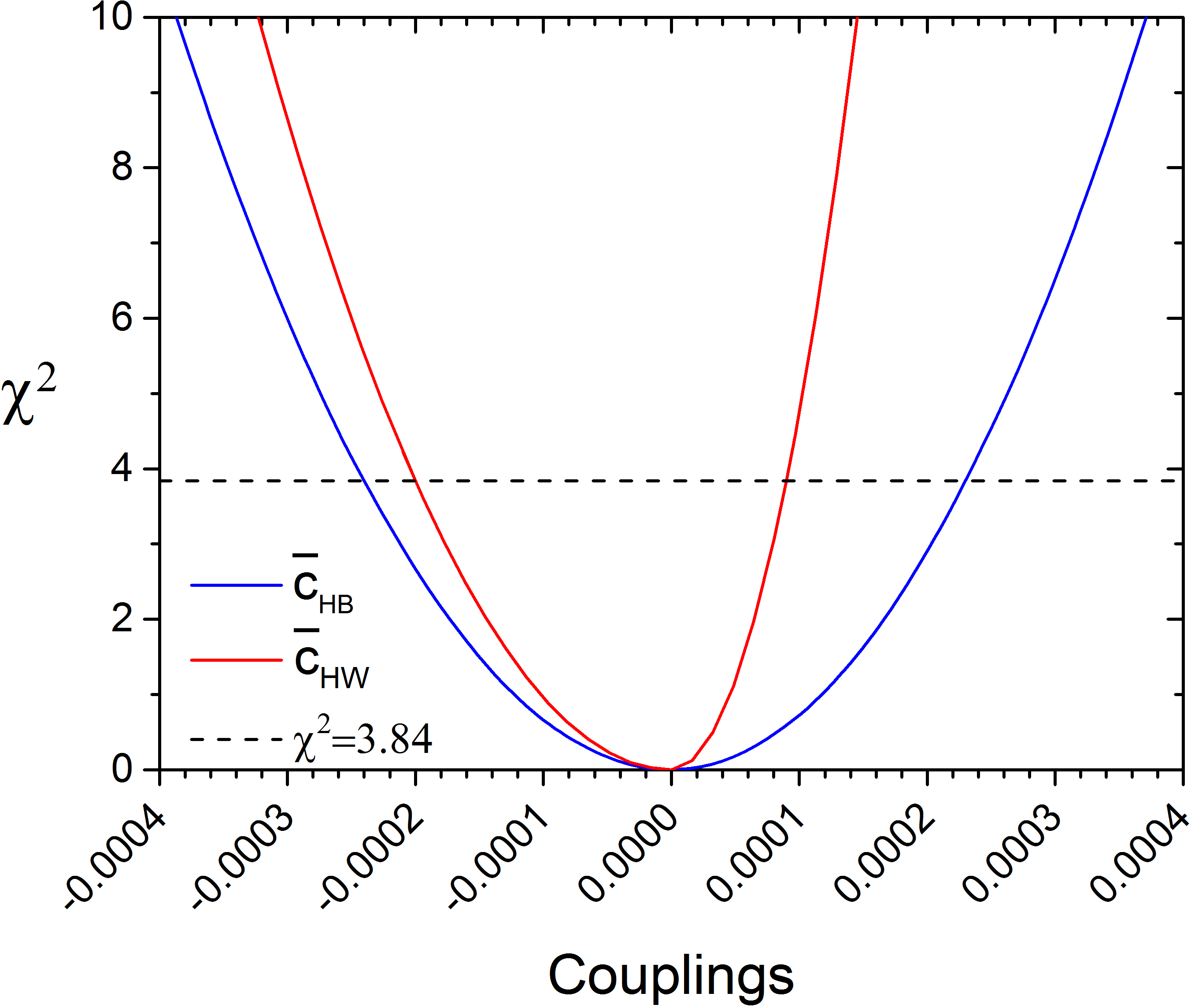}
\caption{}
\label{fig11:b}
\end{subfigure}\hfill
\caption{Obtained $\chi^2$ with one-parameter analysis as function of $\overline{c}_{HB}$ and $\overline{c}_{HW}$ coefficients at CLIC (a) and Muon Collider (b). Dotted line corresponds to 95\% C.L.}
\label{fig11}
\end{figure}

In this study, we focus on obtaining the 95\% C.L. limits of $\overline{c}_{HB}$ and $\overline{c}_{HW}$ coefficients through the $\ell^- \ell^+ \rightarrow HZZ$ signal process with a stage at $\sqrt{s}=3$ TeV and ${\cal L}_{\text{int}}=5$ ab$^{-1}$ for CLIC and a stage at $\sqrt{s}=10$ TeV and ${\cal L}_{\text{int}}=10$ ab$^{-1}$ for Muon Collider. Fig.~\ref{fig11} shows the obtained $\chi^2$ value as a function of coefficients $\overline{c}_{HB}$ and $\overline{c}_{HW}$ after applying all cuts with 90\% b-tagging efficiency at CLIC and Muon Collider. At the same time, the dotted line at $\chi^2=3.84$ corresponds to 95\% C.L. The intersections of the $\chi^2$ curve with this dotted line give the 95\% C.L. limits.

The 95\% C.L. limits of $\overline{c}_{HB}$ and $\overline{c}_{HW}$ coefficients, including detector effects, are given in Table~\ref{tab4} using the one-parameter analysis. Since $H \rightarrow b \bar{b}$ is the decay channel in the final state of the signal process, b-tagging efficiency has a significant role in the analysis. This effect is demonstrated in Table~\ref{tab4} through a comparison of b-tagging efficiencies for CLIC and Muon Collider at two working points. According to the results in the table, it is emphasized that the sensitivity of the limits for CLIC and Muon Collider increases with increasing b-tagging efficiency, and a better limit is reached with the loose working point (90\% b-tagging efficiency). Furthermore, comparing the limits of recent experimental and previous phenomenological studies with our own limits, as shown in Table~\ref{tab4}, is one of the notable aims of our study. In addition to the limits obtained using the formulation in which the effective scale $\Lambda$ of the SILH Lagrangian is defined by the mass of the $W$-boson in $m_W$, the limits rescaled with $\Lambda=1$ TeV are given in Table~\ref{tab4} with a scaling factor of (1 TeV/$m_W$)$^2 \approx 154.818$ for the reader to sense. A detailed comparison of the limits in Table~\ref{tab4} will be presented in Section~\ref{Sec5}.

\begin{table}[H]
\centering
\caption{The 95\% C.L. limits on $\overline{c}_{HB}$ and $\overline{c}_{HW}$ coefficients in various experimental and phenomenological studies, and in our study.}
\label{tab4}
\begin{tabular}{p{2cm}p{1.1cm}p{3.3cm}p{3.7cm}p{3cm}p{3cm}}
\hline
\hline
\multicolumn{2}{l}{Coefficients} & \centering $\overline{c}_{HB}$ & \centering $\overline{c}_{HW}$ &&\\
\hline
\multirow{2}{*}{Experimental} &
& $[-0.057;0.051]$ &  $[-0.057;0.051]$ \cite{Aaboud:2018yer} &&\\
\multirow{2}{*}{Results} & & $[-0.090;0.032]$ &  $[-0.005;0.006]$ \cite{Aaboud:2019ygw} &&\\
& & $[-0.025;0.022]$ & $[-0.025;0.022]$ \cite{ATLAS:2019sdf} &&\\
\hline

& & $[-0.0172; 0.00661]$ & $[-0.00187; 0.0018]$ \cite{Khanpour:2017opf} &&\\
\multirow{2}{*}{Phenomenological}
& & $[-0.39; 0.41]$ & $[-0.18; 0.17]$ \cite{Kumar:2019twv} &&\\
 \multirow{2}{*}{Results} & & $[-0.0482; 0.0153]$ & $[-0.00658; 0.00555]$ \cite{Denizli:2018rca} &&\\
& & $[-2.5608; 1.3256]$ & $[-0.7692; 0.3969]$ \cite{Spor:2024ghq} &&\\
& & $[-0.0058; 0.0057]$ & $[-0.0047; 0.0035]$ \cite{Spor:2025eaz} &&\\
\hline
\underline{This study} & b-tag. & \centering $\overline{c}_{HB}$ & \centering $\overline{c}_{HW}$ & $\overline{c}_{HB}$ ($\Lambda = 1$ TeV) & $\overline{c}_{HW}$ ($\Lambda = 1$ TeV)\\ 
\multirow{2}{*}{CLIC} & 70\% & $[-0.00140;0.00098]$ & $[-0.00172;0.00028]$ & $[-0.217;0.152]$ & $[-0.266;0.043]$\\
& 90\% & $[-0.00138;0.00090]$ & $[-0.00162;0.00026]$ & $[-0.214;0.139]$ & $[-0.251;0.040]$\\ \cline{2-6}
Muon & 70\% & $[-0.00029;0.00026]$ & $[-0.00024;0.00013]$ & $[-0.045;0.040]$ & $[-0.037;0.020]$\\
Collider & 90\% & $[-0.00024;0.00023]$ & $[-0.00020;0.00009]$ & $[-0.037;0.036]$ & $[-0.031;0.014]$\\
\hline \hline
\end{tabular}
\end{table}

\begin{figure}[H]
\centering
\begin{subfigure}{0.5\linewidth}
\includegraphics[width=\linewidth]{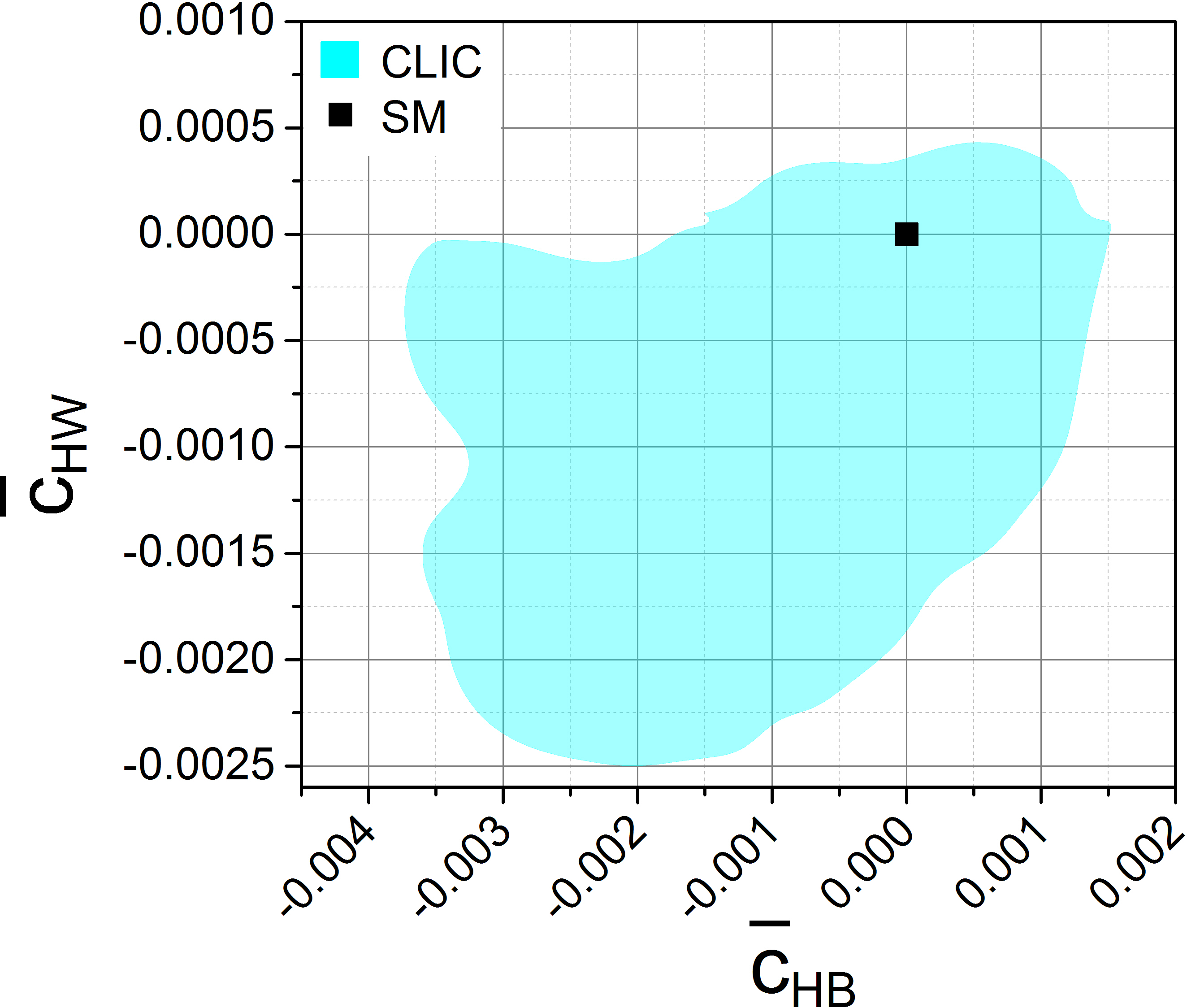}
\caption{}
\label{fig12:a}
\end{subfigure}\hfill
\begin{subfigure}{0.5\linewidth}
\includegraphics[width=\linewidth]{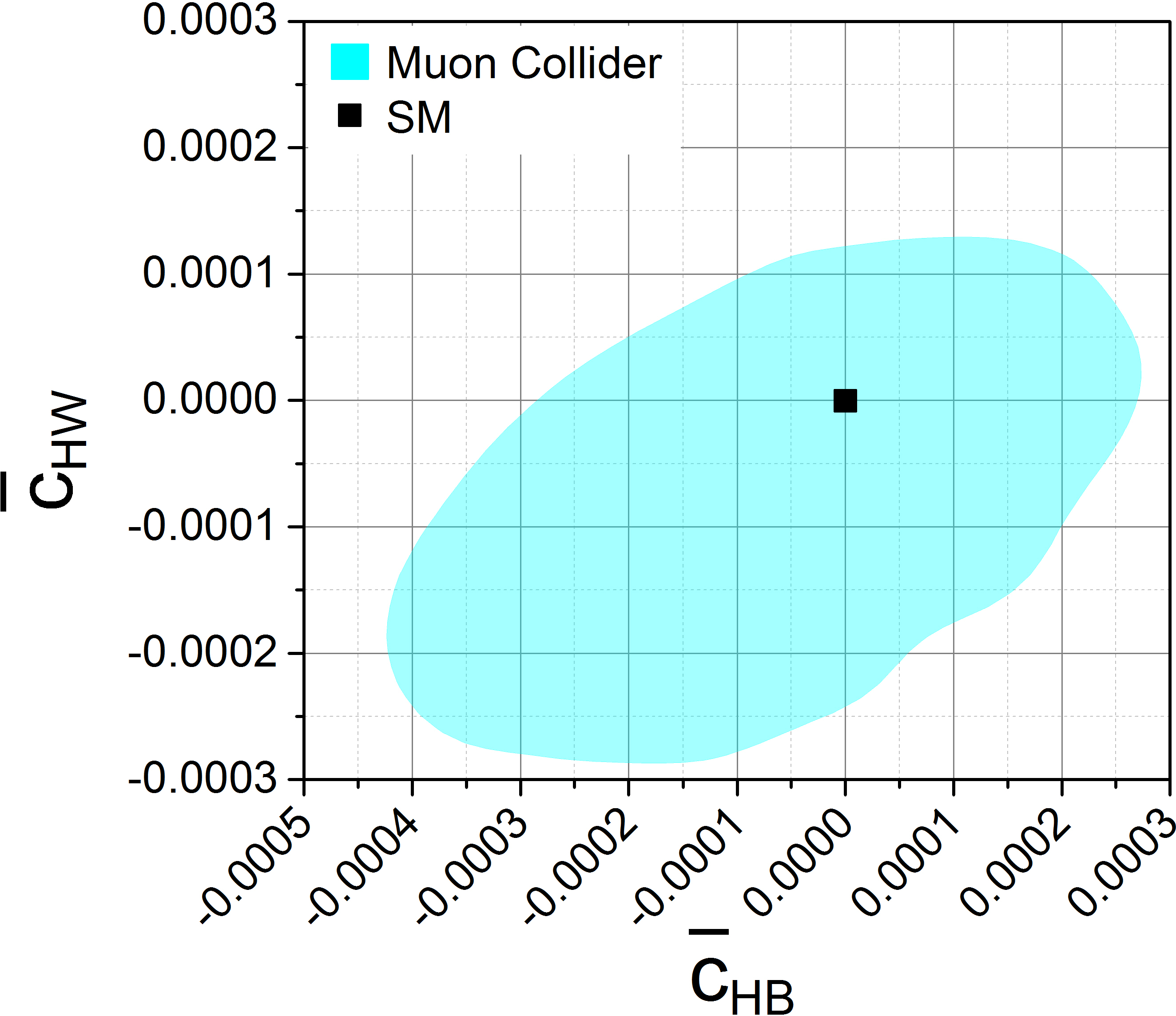}
\caption{}
\label{fig12:b}
\end{subfigure}\hfill
\caption{Two-dimensional 95\% C.L. intervals in $\overline{c}_{HB}-\overline{c}_{HW}$ plane at CLIC (a) and Muon Collider (b).}
\label{fig12}
\end{figure}

The 95\% C.L. contour for examining the relationship between the two coefficients in the $\overline{c}_{HB}-\overline{c}_{HW}$ plane at CLIC and Muon Collider using two-parameter analysis is given in Fig.~\ref{fig12}. Furthermore, in Fig.~\ref{fig12}, the contours in both colliders are evaluated for 90\% b-tagging efficiency. In this analysis, the 95\% C.L. contour is obtained with a $\chi^2$ value of $5.99$ corresponding to two degrees of freedom.

\section{Conclusions} \label{Sec5}

Among future colliders, CLIC and Muon Collider are attractive options recently due to their multi-TeV energies and clean environments for exploring Higgs interactions in new physics beyond the SM. We investigate the dimension-six operators of anomalous Higgs-gauge boson couplings $H\gamma Z$ and $HZZ$ via the process $\ell^-\ell^+ \rightarrow HZZ$ at CLIC and Muon Collider using an effective Lagrangian approximation in the SILH basis. We analyze the production of $bb\ell \ell \nu\nu$ final state along with the decay of $H$ and $Z$-bosons, considering the realistic detector effects. We present the kinematic distributions of the signal and relevant background processes using cut-based analysis to separate signal events from backgrounds. The 95\% C.L. limits on the Wilson coefficients $\overline{c}_{HB}$ and $\overline{c}_{HW}$ in the SILH basis are obtained using the $\chi^2$ test from the invariant mass distribution of the $\ell \ell bb$ system for two working points of b-tagging efficiency. The best limits in our study are obtained at the loose working point (90\% b-tagging efficiency); these are $[-0.00138; 0.00090]$ for the coefficient $\overline{c}_{HB}$ and $[-0.00162; 0.00026]$ for the coefficient $\overline{c}_{HW}$ at CLIC, and $[-0.00024; 0.00023]$ for the coefficient $\overline{c}_{HB}$ and $[-0.00020; 0.00009]$ for the coefficient $\overline{c}_{HW}$ at Muon Collider.

In Table~\ref{tab4}, we present a comparison between the 95\% C.L. limits of the $\overline{c}_{HB}$ and $\overline{c}_{HW}$ coefficients from this study and those reported in various phenomenological and experimental studies. First, let us consider the comparison of the sensitivities of $\overline{c}_{HB}$ and $\overline{c}_{HW}$ coefficients at CLIC and Muon Collider in this study with experimental studies. Among the experimental studies, the most sensitive limit at 95\% C.L. for the coefficient $\overline{c}_{HW}$ was determined by ATLAS collaboration \cite{Aaboud:2019ygw} as $[-0.005; 0.006]$ at $\sqrt{s}=13$ TeV with ${\cal L}_{\text{int}}=79.8$ fb$^{-1}$, and the most sensitive limit at 95\% C.L. for the coefficient $\overline{c}_{HB}$ was obtained by ATLAS collaboration \cite{ATLAS:2019sdf} as [-0.025; 0.022] at $\sqrt{s}=13$ TeV with ${\cal L}_{\text{int}}=139$ fb$^{-1}$. The 95\% C.L. limits of $\overline{c}_{HB}$ coefficient at CLIC and Muon Collider in our study are approximately 21 times and 100 times more sensitive than the limit in Ref.~\cite{ATLAS:2019sdf}, respectively, and the 95\% C.L. limits of $\overline{c}_{HW}$ coefficient at CLIC and Muon Collider are approximately 6 times and 38 times more sensitive than the limit in Ref.~\cite{Aaboud:2019ygw}, respectively.

Table~\ref{tab4} presents the limits of some phenomenological studies involving different processes and decay channels at ILC \cite{Khanpour:2017opf,Kumar:2019twv}, CLIC \cite{Denizli:2018rca}, and Muon Collider \cite{Spor:2024ghq,Spor:2025eaz}. While we employ the SILH basis in this study, model-independent limits on the triple Higgs coupling at the 500 GeV ILC have been investigated using a different operator basis \cite{Barklow:2018yvb,Barklow:2018muv}. The most sensitive limit for the coefficient $\overline{c}_{HB}$ was given as $[-0.0058; 0.0057]$ in Ref.~\cite{Spor:2025eaz} performed at Muon Collider with $\sqrt{s}=10$ TeV and ${\cal L}_{\text{int}}=10$ ab$^{-1}$. The most sensitive limit for the coefficient $\overline{c}_{HW}$ was obtained as $[-0.00187; 0.0018]$ in Ref.~\cite{Khanpour:2017opf} performed at an electron-positron collider with $\sqrt{s}=350$ GeV and ${\cal L}_{\text{int}}=3000$ fb$^{-1}$. In our study, the 95\% C.L. limits of the $\overline{c}_{HB}$ coefficient at CLIC and Muon Collider are approximately 5 times and 24 times more sensitive than the limit in Ref.~\cite{Spor:2025eaz}, respectively. The 95\% C.L. limits of the $\overline{c}_{HW}$ coefficient at CLIC and Muon Collider are approximately 2 times and 13 times more sensitive than the limit in Ref.~\cite{Khanpour:2017opf}, respectively.

The clean experimental environment of the future lepton colliders, with their high center-of-mass energy and integrated luminosity, presents a very promising option for testing the SM predictions and searching new physics beyond the SM. The fact that the limits on $\overline{c}_{HB}$ and $\overline{c}_{HW}$ coefficients for anomalous $H\gamma Z$ and $HZZ$ couplings at CLIC and Muon Collider are more sensitive than those from experimental studies at LHC highlights the potential of CLIC and Muon Collider. The study of Higgs physics will be greatly enriched by the combined insights from future lepton colliders and supplementary measurements from hadron colliders. The results certainly encourage a more detailed and comprehensive investigation of lepton colliders.

\section*{Declaration of competing interest}

The authors declare that they have no known competing financial interests or personal relationships that could have appeared
to influence the work reported in this paper.

\end{document}